\documentclass[12pt,a4]{article}
\usepackage{graphicx,subfigure}
\usepackage{amsmath}
\usepackage{amssymb}
\usepackage{bm}
\begin{document}

\title{
\bf \large \bf
Cross-fertilization of Ferreira's Hopfions And Electromagnetic Knots\\[30pt]}
\author{ Chang-Guang Shi $^{1}$ and Minoru Hirayama$^{2}$\\[10pt]
\\{\normalsize  {\sl $^{1}$Department of Mathematics and Physics}}
\\{\normalsize {\sl
Shanghai University of Electric Power }}
\\{\normalsize {\sl Pinglian Road 2103, Shanghai 200090, China}}
\\{\normalsize  {\sl $^{2}$Department of Physics, University of Toyama }}
\\{\normalsize  {\sl Gofuku 3190, Toyama 930-8555, Japan}}
\\{\normalsize{\sl $^1$Email  : shichangguang@shiep.edu.cn}}
\\{\normalsize {\sl $^2$Email  : hirayama@jodo.sci.u-toyama.ac.jp}}}

\date{December 2008}

\maketitle
\begin{abstract}
The interrelation between Ferreira's Hopf solitons of a conformal
nonlinear $\sigma$ model and the electromagnetic knots found by
Ra$\tilde{\rm{n}}$ada et al. is investigated. It is shown that the
electromagnetic knots yield exact
 solutions of the conformal nonlinear $\sigma$ model
  different from those obtained by Ferreira.
Conversely,  It is discussed that Ferreira's solutions realize magnetic knots.
The energy associated with these two kinds of knots are compared.
The structure of the electric charge distribution and the electric current density
 associated with the magnetic knots is investigated.
\end{abstract}

\vspace{1cm}
---------------------------------------------------------------------------------------------

\small{Corresponding author: Chang-Guang Shi,
shichangguang@shiep.edu.cn}

\newpage
\section{Introduction}
Recently, the electromagnetic knots have attracted
 much attention\cite{Trautman,Ranada1,Ranada2}.
 They are the solutions of the Maxwell equations in the vacuum
 possessing knot structures.
As every physicist knows, the Maxwell equations in the vacuum are given by
\begin{eqnarray}
\partial_{\mu}F^{\mu\nu}=0,\quad \partial_{\mu}\tilde{F}^{\mu\nu}=0,
\label{Maxwell}
\end{eqnarray}
where $F_{\mu\nu}$ and $\tilde{F}^{\mu\nu}$ are defined by
\begin{eqnarray}
F_{\mu\nu}=\partial_{\mu}A_{\nu}-\partial_{\nu}A_{\mu},\quad
\tilde{F}^{\mu\nu}=\frac{1}{2}\varepsilon^{\mu\nu\alpha\beta}F_{\alpha\beta}
\end{eqnarray}
in terms of the $4$-potential $A_{\mu}$.
Here $\varepsilon^{\mu\nu\alpha\beta}$
is the totally anti-symmetric Levi-Civita tensor.
They might be important in plasma physics and fluid dynamics.
 Besides theoretical interests, it was discussed that the electromagnetic
  knots might be the origin of the phenomenon of
  ball lightning\cite{Ranada-Trueba,Ranada-Soler-Trueba,IB}.\\
On the other hand, Ferreira\cite{Fer} succeeded in obtaining the $3+1$
dimensional solutions of a model,
which we refer to as the conformal nonlinear $\sigma$ model (CNLSM) in this paper,
 for a complex scalar field.
   It is expected that the CNLSM
  has  connections to the low energy limit of the Yang-Mills
 theory and the Skyrme-Faddeev model\cite{Fad,FN}.\\
 Because of the conformal symmetry of the electromagnetism and the CNLSM, the
 solutions of both theories can involve a parameter specifying the
 space-time scale
 and the energies associated with these solutions
 are proportional to the inverse of this scale parameter.
Therefore they  cannot be energetically-stable configurations.
The solutions of both theories, however, can be topologically-stable
 in the sense that conserved topological numbers
 can be defined for them.\\
 Two Hopf indices are defined for an electromagnetic knot,
 while a single Hopf index is defined for a solution of the CNLSM.\\
 The Lagrangian density of the CNLSM is given by\cite{Fer}
\begin{equation}
\mathcal{L}_F=-\frac{1}{4}H_{\mu\nu}H^{\mu\nu},
\end{equation}
where $H_{\mu\nu}$ is defined by
\begin{equation}
H_{\mu\nu}=\frac{1}{2}\bm{n}\cdot(\partial_{\mu}\bm{n}\times \partial_{\nu}\bm{n})
=\frac{1}{i}\frac{(\partial_{\mu} u  \partial_{\nu} u^*-\partial_{\nu} u \partial_{\mu} u^*)}
{(1+u u^*)^2}.
\label{Fer.Def}
\end{equation}
Here $\bm{n}$ denotes a point on $S^2$
\begin{equation}
\boldsymbol{n}=\left(n^1,n^2,n^3\right),\quad {\boldsymbol{n}}^2
=\boldsymbol{n}\cdot\boldsymbol{n}
=\sum\limits_{a=1}^{3}n^a n^a=1
\end{equation}
which is related to the complex field $u$ by the stereographic projection
\begin{equation}
\bm{n}=\frac{1}{1+u u^*}\left(u+u^*, -i(u-u^*),
u u^*-1\right)\quad {\rm{or}}\quad u=\frac{n_1+in_2}{1-n_3}.
\end{equation}
Regarding $u$ and $u^*$ as the fundamental fields, the field equations are given by
\begin{equation}
\partial_{\mu}\left( H^{\mu\nu}\partial_{\nu}u \right )=0,\quad
\partial_{\mu}\left( H^{\mu\nu}\partial_{\nu}u^* \right )=0. \label{Fer.FE}
\end{equation}

In this paper, we discuss that these two theories can cross-fertilise each other.
We show that the electromagnetic configurations discussed in the theory of
 electromagnetic knots supplies us with a new class of solutions of the CNLSM different
from those obtained by Ferreira.

 Conversely, we show also that
Ferreira's solution of the CNLSM, which we hereafter refer to as
F-Hopfions, supplies us with a class of exact magnetic knot
configurations in some
  electric charge and current distributions.
  We investigate the electric field, magnetic field, electromagnetic energy,
  electric charge density and the electric current density which
  arise from F-Hopfions.
  If we adopt the same scale parameter in the simplest nontrivial examples
  of the two theories, it turns out
  that the energy of the electromagnetic field obtained from F-Hopfion
   is equal to the half of that of the configuration
   discussed in the theory of electromagnetic knot.

     This paper is organized as follows.
In Sec.2, we first discuss that F-Hopfions are the $3+1$-dimensional
generalizations of the Hopf fibration.
We then  compare the simplest F-Hopfion with
the configuration of the complex scalar fields appearing in
the electromagnetic knots. We next obtain new solutions of the
CNLSM which are different from F-Hopfions.
In Sect.3, we investigate the electromagnetism implied by
F-Hopfions. It corresponds to the electromagnetism
not in the vacuum but in non-vanishing  electric charge and current  distributions.
We discuss some properties of them.
The final section is devoted to summary.

\section{F-Hopfion and Hopf fibration}
Guided by the invariance of CNLSM under the conformal group $SO(4,2)$ of
the four-dimensional Minkowski space-time\cite{Bab-Fer}, the following
variables $Y, \zeta, \varphi$ and $\xi$ are introduced in \cite{Fer}.
Expressing $(x^0,x^1,x^2,x^3)$ as $(t,x,y,z)$, they are defined by
\begin{eqnarray}
&\displaystyle{t=\frac{a}{p} \sin \zeta,\quad x= \frac{a}{p} \frac{ \cos \varphi}{\sqrt{1+Y}}},\quad
\displaystyle{y=\frac{a}{p} \frac{ \sin \varphi}{\sqrt{1+Y}}},
\quad \displaystyle{z=\frac{a}{p} \sin \xi \sqrt{\frac{Y}{1+Y}}},\nonumber \\
&\displaystyle{p=\cos \zeta-\cos \xi \sqrt{\frac{Y}{1+Y}},}
\end{eqnarray}
where $a$ is an arbitrary constant parameter of the dimension of length which fixes
the space-time scale.
For simplicity we hereafter set  $a=1$. Then we have
\begin{eqnarray}
&\displaystyle{Y=\frac{\left(1+s^2  \right)^2+4z^2}{4\rho^2},\quad \tan\varphi=\frac{y}{x}},\quad
\displaystyle{\tan\zeta=\frac{2t}{1-s^2}},\quad \displaystyle{\tan \xi=-\frac{2z}{1+s^2}.}\nonumber\\
&\displaystyle{s^2=t^2-r^2,\quad r^2=\rho^2+z^2,\quad \rho^2=x^2+y^2.}
\end{eqnarray}
It is striking that the Ansatz
\begin{eqnarray}
u=\sqrt{\frac{1-g}{g}}{\rm{e}}^{i\Phi},\quad g=g(Y),
\quad \Phi=\Phi(\xi, \varphi, \zeta)=m_1\xi+m_2\varphi+m_3\zeta,\label{Fer.An}
\end{eqnarray}
with $m_1, m_2, m_3$ being integers and $m_1+m_2+m_3$ being an even integer
is compatible with the field equation and the
 single-valuedness of $u$ \cite{Fer}. \\
  Under the Ansatz (\ref{Fer.An}), the field equation (\ref{Fer.FE})
 is reduced to a linear ordinary
  differential equation
\begin{equation}
\frac{d}{dY}\left(\Lambda \frac{dg}{dY}\right)=0,\quad \Lambda=m_1^2(1+Y)+m_2^2Y(1+Y)-m_3^2Y.
\end{equation}
The solutions of this equation can be classified by the parameters
$\Delta$ and $b$ defined by $ \Delta
=\frac{1}{4m_2^2}\left[(m_1+m_3)^2-m_2^2\right]\left[(m_1-m_3)^2-m_2^2\right],
b=\frac{m_1^2+m_2^2-m_3^2}{2m_2^2} $ and the Hopf index of the
solution was calculated  to be $Q_H=m_1m_2[g(0)-g(\infty)]$
\cite{Fer}.

We first discuss the simplest nontrivial case ($m_1=m_2=1,~m_3=0$) briefly.
In this case, the field equation becomes
 $ \frac{d}{dY}\left[(Y+1)^2\frac{dg}{dY}\right]=0$.
 If we adopt the boundary condition
 $g(0)=0, g(\infty)=1$, we obtain $g(Y)=\frac{Y}{Y+1}$, from which we have
 \begin{eqnarray}
 u=\frac{{\rm{e}}^{i(\xi+\varphi)}}{\sqrt{Y}}.
 \end{eqnarray}
 We then find
 \begin{eqnarray}
 u=i\phi_H\quad {\rm{at}}\quad t=0,\label{phiH}
 \end{eqnarray}
 where $\phi_H$ is the Hopf fibration
 \begin{eqnarray}
 \phi_H=\frac{2(x+iy)}{2z+i(r^2-1)}.
 \end{eqnarray}
In other words, the simplest example of F-Hopfion is a
$3+1$-dimensional generalization of the Hopf fibration.
We note that
the Hopf index of the above $u$ is equal\vspace{5mm} to $-1$.\\
The $3+1$-dimensional generalizations of $\phi_H$ different from
the above $u$ are given by \cite{Ranada1,Ranada2,Ranada-Trueba,IB}
\begin{eqnarray}
&\displaystyle{\eta_m=-\frac{[Ky+t(K-1)]+i(tz-Kx)}{(Kz+tx)+i[K(K-1)-ty]}},\nonumber\\
&\displaystyle{ \eta_e=i\frac{(Kz+tx)+i[Ky+t(K-1)]}{(tz-Kx)+i[K(K-1)-ty]},}\nonumber\\
&K\displaystyle{\equiv\frac{1}{2}(1-s^2)=\frac{1}{2}(r^2-t^2+1)}.\label{eta}
\end{eqnarray}
Although we can introduce
  the scale parameter $a$ similarly to the case of F-Hopfion,
   we consider the case of $a=1$ again for simplicity.
   The remarkable property of the pair ($\eta_e$, $\eta_m$) is  that  they satisfy
\begin{eqnarray}
\displaystyle{\frac{1}{i(1+\eta_m\eta_m^*)^2}\frac{\partial(\eta_m, \eta_m^*)}{\partial(y,z)}
=\frac{1}{i(1+\eta_e\eta_e^*)^2}\frac{\partial(\eta_e, \eta_e^*)}{\partial(t,x)},} \nonumber
\end{eqnarray}
\begin{eqnarray}
\displaystyle{\frac{1}{i(1+\eta_e\eta_e^*)^2}\frac{\partial(\eta_e, \eta_e^*)}{\partial(y,z)}
=-\frac{1}{i(1+\eta_m\eta_m^*)^2}\frac{\partial(\eta_m, \eta_m^*)}{\partial(t,x)}}.
\label{duality}
\end{eqnarray}
We also have the relations obtained by replacing $\{(y,z),(t,x)\}$ in (\ref{duality}) by
$\{(z,x),(t,y)\}$ and $\{(x,y),(t,z)\}$.
It is easy to find
\begin{eqnarray}
\eta_m=i\phi_H,\quad \eta_e=\left(i\phi_H\right)_{(x,y,z)\rightarrow(z,y,-x)}\quad {\rm{at}}~ t=0. \label{phiHH}
\end{eqnarray}
They were the starting configurations of the discussion
 of the electromagnetic knots  \cite{Ranada1,Ranada2}.\\
 Comparing (\ref{phiHH}) with (\ref{phiH}), we see that the simplest F-Hopfion coincides with
 $\eta_m$ at $t=0$.
  Rewriting $\eta_m$ and $\eta_e$ in terms of $Y, \zeta, \varphi, \xi$,
we have
\begin{eqnarray}
& \displaystyle{\eta_m=-\frac{\rm{e}^{i\varphi}- \sqrt{Y}\rm{e}^{i\xi}\tan \zeta}
{\sqrt{Y}\rm{e}^{-i\xi}+i\rm{e}^{-i\varphi}\tan\zeta}},\\
& \displaystyle{\eta_e=-\left(\frac{\rm{e}^{i\varphi}- \sqrt{Y}\rm{e}^{i\xi}\tan \zeta}
{\sqrt{Y}\rm{e}^{-i\xi}+i\rm{e}^{-i\varphi}\tan\zeta}\right)_{(x,y,z)\rightarrow(z,y,-x)}}.
\end{eqnarray}
With the help of (\ref{duality}) and the others, it can be readily seen that
both $\eta_m$ and $\eta_e$ also
   solve the CNLSM.
It is clear that neither $\eta_m$ nor $\eta_e$ satisfy Ferreira's
 Ansatz (\ref{Fer.An}).
Thus we have found new solutions of the CNLSM from the configurations
 found in the theory
of the electromagnetic knots.
\section{Electromagnetism implied by F-Hopfion}
As for the electromagnetic knots, the following elctric field $\bm{E}$ and the magnetic field $\bm{B}$
were discussed in \cite{IB}:
\begin{eqnarray}
\boldsymbol{E}=\frac{1}{i}\frac{\nabla\eta_e\times\nabla\eta_e^*}{(1+\eta_e \eta_e^*)^2},
\quad \boldsymbol{B}=\frac{1}{i}\frac{\nabla\eta_m\times\nabla\eta_m^*}
{(1+\eta_m \eta_m^*)^2}\label{EB}.
\end{eqnarray}
From the definition (\ref{eta}) and the property (\ref{duality}),
$\bm{E}$ and $\bm{B}$ satisfy the Maxwell equations in the vacuum :
\begin{eqnarray}
&\displaystyle{{\nabla\cdot \boldsymbol{B}}=0,}\quad
 \displaystyle{\frac{\partial {\boldsymbol{B}}}{\partial t}+{\nabla\times\boldsymbol{E}}=0,}\nonumber\\
&\displaystyle{{\nabla\cdot\boldsymbol{E}}=0},\quad
\displaystyle{\frac{\partial{\boldsymbol{E}}}{\partial t}-\nabla\times\boldsymbol{B}}
\end{eqnarray}
and the constraint
\begin{eqnarray}
&\bm{E}\cdot\bm{B}=0.
\end{eqnarray}

We now define the electric and magnetic fields associated with Ferreira's solution by
\begin{eqnarray}
&\boldsymbol{B}_F=
\left(B_{F,x},B_{F,y},B_{F,z}\right)=-\left(H_{23},H_{31},H_{12}\right),\nonumber \\
&\boldsymbol{E}_F=
\left(E_{F,x},E_{F,y},E_{F,z}\right)=\left(H_{01},H_{02},H_{03}\right).
\end{eqnarray}
From the definition (\ref{Fer.Def}), we obtain
$\varepsilon^{\mu\nu\alpha\beta}\partial_{\nu}H_{\alpha\beta}=0$
 which is equivalent to
\begin{equation}
{\nabla\cdot \boldsymbol{B}_F}=0,\quad
 \frac{\partial {\boldsymbol{B}_F}}{\partial t}+{\nabla\times\boldsymbol{E}_F}=0. \label{Eq1}
\end{equation}
The definition (\ref{Fer.Def}) also yields
\begin{equation}
{\boldsymbol{E}_F\cdot\boldsymbol{B}_F}=0.
\end{equation}
The field equation (\ref{Fer.FE})
yields the constraints
\begin{eqnarray}
\rho_F\frac{\partial Y}{\partial t}+\boldsymbol{j}_F\cdot \nabla Y=0,\quad
\rho_F\frac{\partial \Phi}{\partial t}+\boldsymbol{j}_F\cdot \nabla \Phi=0,\label{Fer.Con}
\end{eqnarray}
where $\rho_F$ and $\bm{j}_F$ are defined by
\begin{eqnarray}
{\nabla\cdot\boldsymbol{E}_F}=\rho_F,\quad
\nabla\times
\boldsymbol{B}_F-\frac{\partial{\boldsymbol{E}_F}}{\partial t}
=\boldsymbol{j}_F.\label{Fer.Density}
\end{eqnarray}
Regarding (\ref{Eq1}) and (\ref{Fer.Density}) as the Maxwell equations, $\rho_F$ and $\bm{j}_F$ can be interpreted as the electric charge and current densities, respectively. The definition (\ref{Fer.Density})
ensures the continuity relation
\begin{eqnarray}
\frac{\partial \rho_F}{\partial t}+\nabla\cdot \boldsymbol{j}_F=0.
\end{eqnarray}
From the constraint (\ref{Fer.Con}),
we obtain
 \begin{eqnarray}
 {\boldsymbol{E}_F\cdot\boldsymbol{j}_F}=0,\label{Ej}
 \end{eqnarray}
 which guarantees that $\varepsilon_F$ defined by
\begin{eqnarray}
\varepsilon_F\equiv
=\frac{1}{2}\int\!\!\!\int\!\!\!\int dxdydz \left(\bm{E}_F^2+\bm{B}_F^2 \right) \label{Fer.En}
\end{eqnarray}
is conserved.
In \cite{Fer}, it was shown that there exist infinite number of conserved
 quantities in the CNLSM and how to construct them.
  We also find
 \begin{eqnarray}
 {\boldsymbol{B}_F\cdot\boldsymbol{j}_F}=\boldsymbol{B}_F\cdot (\nabla\times \boldsymbol{B}_F)
 -\boldsymbol{E}_F\cdot (\nabla\times \boldsymbol{E}_F),\label{Bj}
 \end{eqnarray}
each term on its r.h.s. being of the Chern-Simons type. Hence the integral of
 $ {\boldsymbol{B}_F\cdot\boldsymbol{j}_F}$ is invariant under a wide class of gauge transformations.\\

The components of $\bm{B}_F$ and $\bm{E}_F$ are calculated as
\begin{eqnarray}
& &\displaystyle{B_{F, x}=\frac{C}{\rho^4\Lambda} \nonumber
\left[yLm_1-xz(1-s^2)m_2+2tyzm_3\right]},\\
& &\displaystyle{B_{F, y}=-\frac{C}{\rho^4\Lambda}
\left[xLm_1+yz(1-s^2)m_2+2txzm_3\right],}\nonumber \\
& &\displaystyle{B_{F, z}=-\frac{C}{2\rho^4\Lambda}
\left[(t^2-z^2)^2-\rho^4+2(t^2+z^2)+1  \right]m_2,} \nonumber\\
& &\displaystyle{L=1+t^2+r^2,}
\end{eqnarray}
and
\begin{eqnarray}
& &\displaystyle{E_{F, x}=-\frac{C}{\rho^4\Lambda}
\left[2txzm_1-ty(1+s^2)m_2+x(1+s^2+2z^2)m_3     \right],}\nonumber \\
& &\displaystyle{E_{F, y}=-\frac{C}{\rho^4\Lambda}
\left[2tyzm_1+tx(1+s^2)m_2+y(1+s^2+2z^2)m_3     \right],}\nonumber \\
& &\displaystyle{E_{F, z}=\frac{2C}{\rho^2\Lambda}
\left(tm_1+zm_3  \right)m_2,}
\end{eqnarray}
where $C$ is a constant. In the case that $g$ satisfies the boundary conditions
 $g(0)=1, g(\infty)=0$, it is fixed as
$
C=\frac{(m_1^2+m_2^2-m_3^2)w}
{\ln\left(\frac{1+w}{1-w}\right)},
 w=\frac{\sqrt{\Delta}}{b}.
$\\
It is tedious but straightforward to obtain $\rho_F$ and $\boldsymbol{j}_F=(j_{F, x},j_{F, y},j_{F, z}) $.
They are given by
\begin{eqnarray}
\displaystyle{\left(-\frac{{\rho^4\Lambda^2}}{2C}\right)\rho_F=
Lm_3(Y^2m_2^2-m_1^2)+2tzm_1[(Y+1)^2m_2^2-m_3^2]}
\end{eqnarray}
and
\begin{align}
\left(-\frac{\rho^4\Lambda^2}{4C}\right)j_{F,x}=&Y^2 (txm_2+ym_3)m_2m_3 \nonumber\\
&+ (Y+1)^2(xzm_2-ym_1)m_1m_2 -x(tm_1+zm_3)m_1m_3  ,\nonumber\\
\left(-\frac{\rho^4\Lambda^2}{4C}\right)j_{F,y}=&Y^2(tym_2-xm_3)m_2m_3\nonumber\\
&+(Y+1)^2(xm_1+yzm_2)m_1m_2-y(tm_1+zm_3)m_1m_3, \nonumber\\
\left(-\frac{\rho^4\Lambda^2}{C}\right)j_{F,z}=&4tzm_3(Y^2m_2^2-m_1^2)\nonumber\\
&+\left[4\rho^2Y+1-s^4  \right][(Y+1)^2m_2^2-m_3^2]m_1.
\end{align}

Up to now, we have presented discussions for general $m_1, m_2$ and $m_3$. We now
consider the simplest nontrivial case $(m_1=m_2=1, m_3=0)$. We set the boundary condition
$g(0)=1$ and $g(\infty)=0$. For this F-Hopfion, the Hopf index is equal to $1$  and we have
\begin{eqnarray}
&\displaystyle{||\bm{B}_F||=\frac{\sqrt{Y+1+t^2}}{\rho^3(Y+1)^{2}}},\\
&\displaystyle{||\bm{E}_F||=\frac{2|t|}{\rho^2(Y+1)^{3/2}},}\\
&\displaystyle{\bm{B}_F^2+\bm{E}_F^2=\frac{(1+4t^2\rho^2)(Y+1)+t^2}{\rho^6(Y+1)^4}.}\label{Fer.en}
\end{eqnarray}
In Fig.1$\sim$3, we show the time-development of $||\bm{B}_F||$,
$||\bm{E}_F||$ and $\bm{B}_F^2+\bm{E}_F^2$
 on the plane $z=1$.

\begin{figure}[htb]
  \subfigure[
t=3]{
      \centering
      \includegraphics[width=4cm]{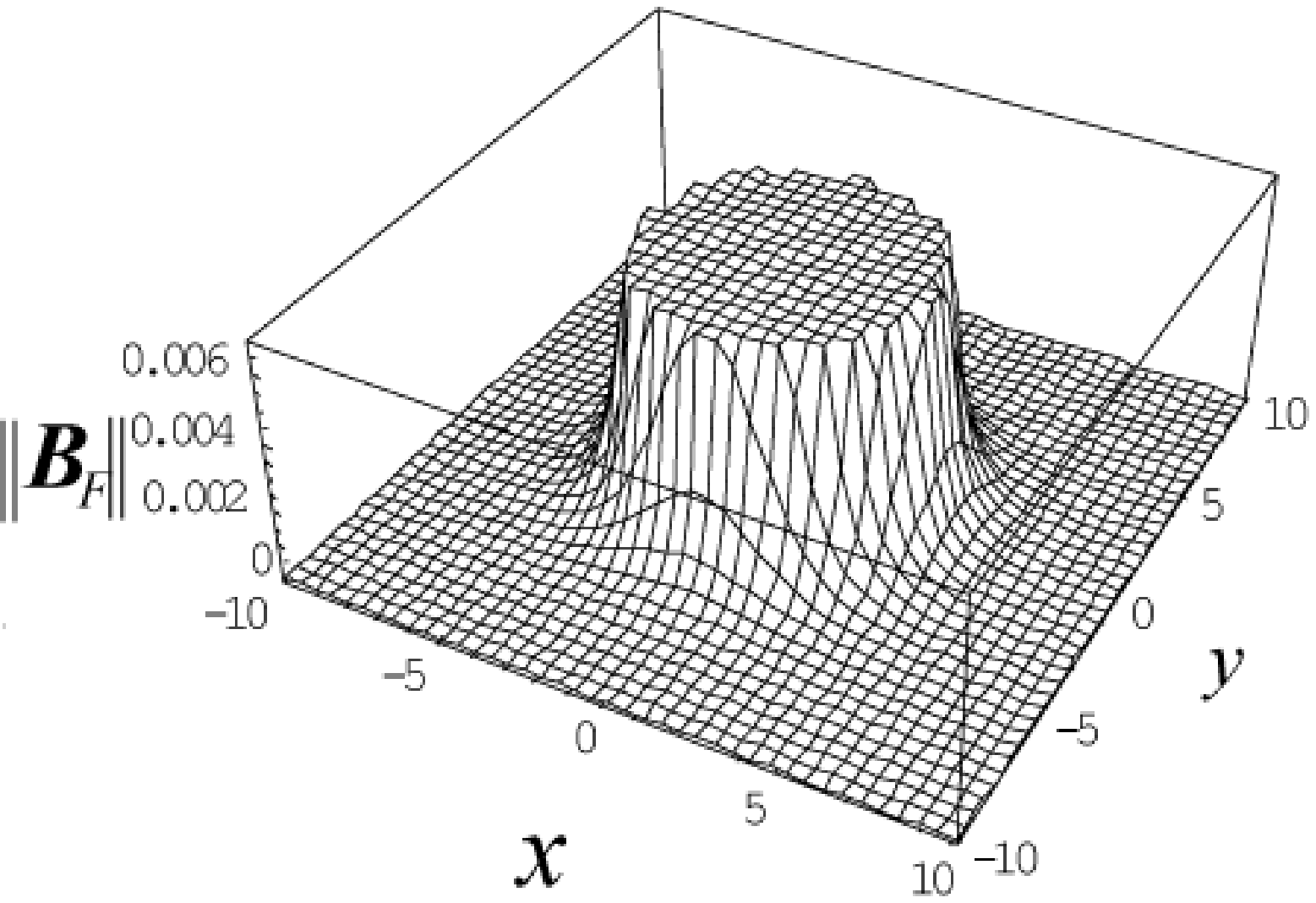}}
  \subfigure[
t=4]{
      \centering
      \includegraphics[width=4cm]{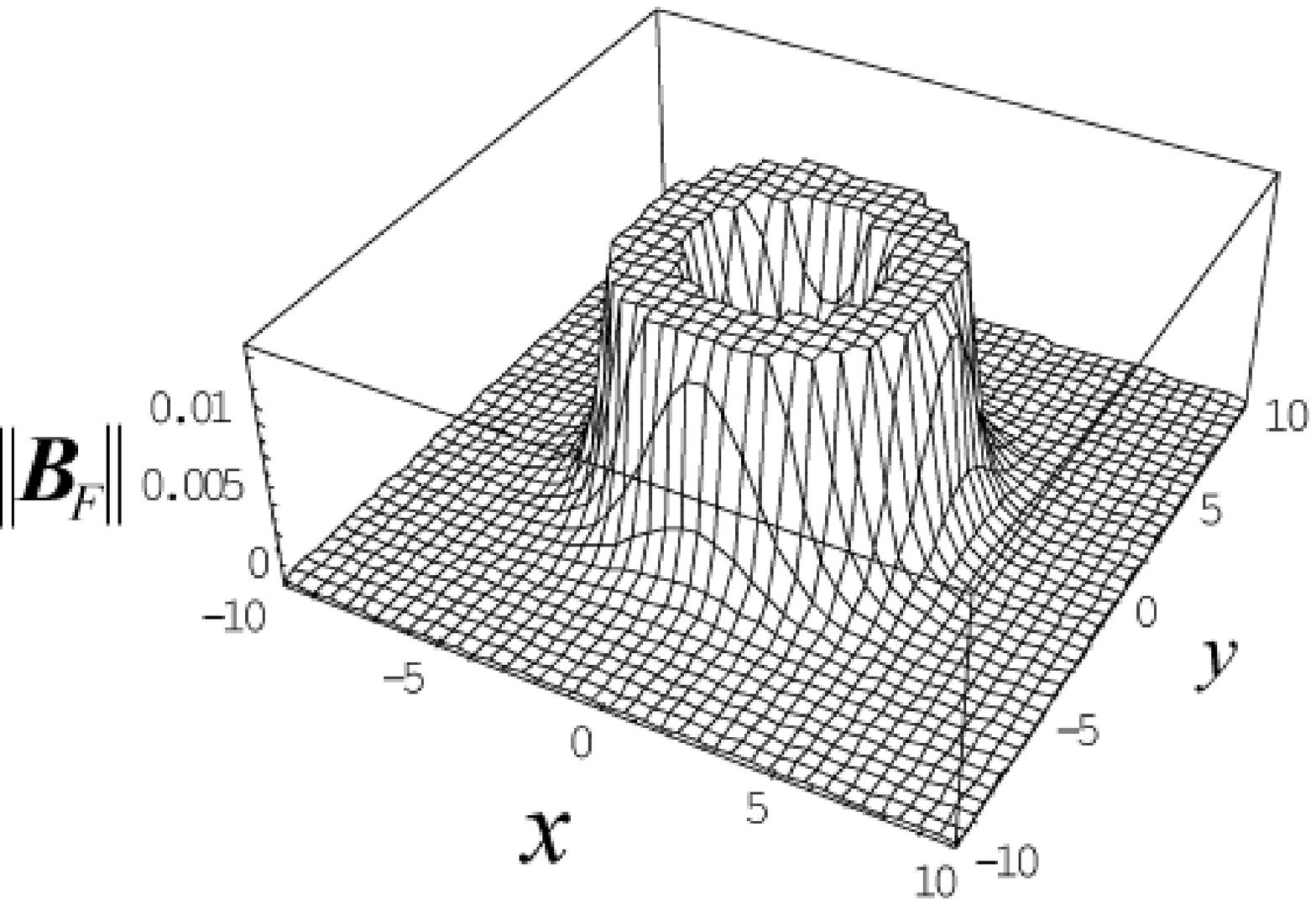}}
       \subfigure[ t=5]{
      \centering
      \includegraphics[width=4cm]{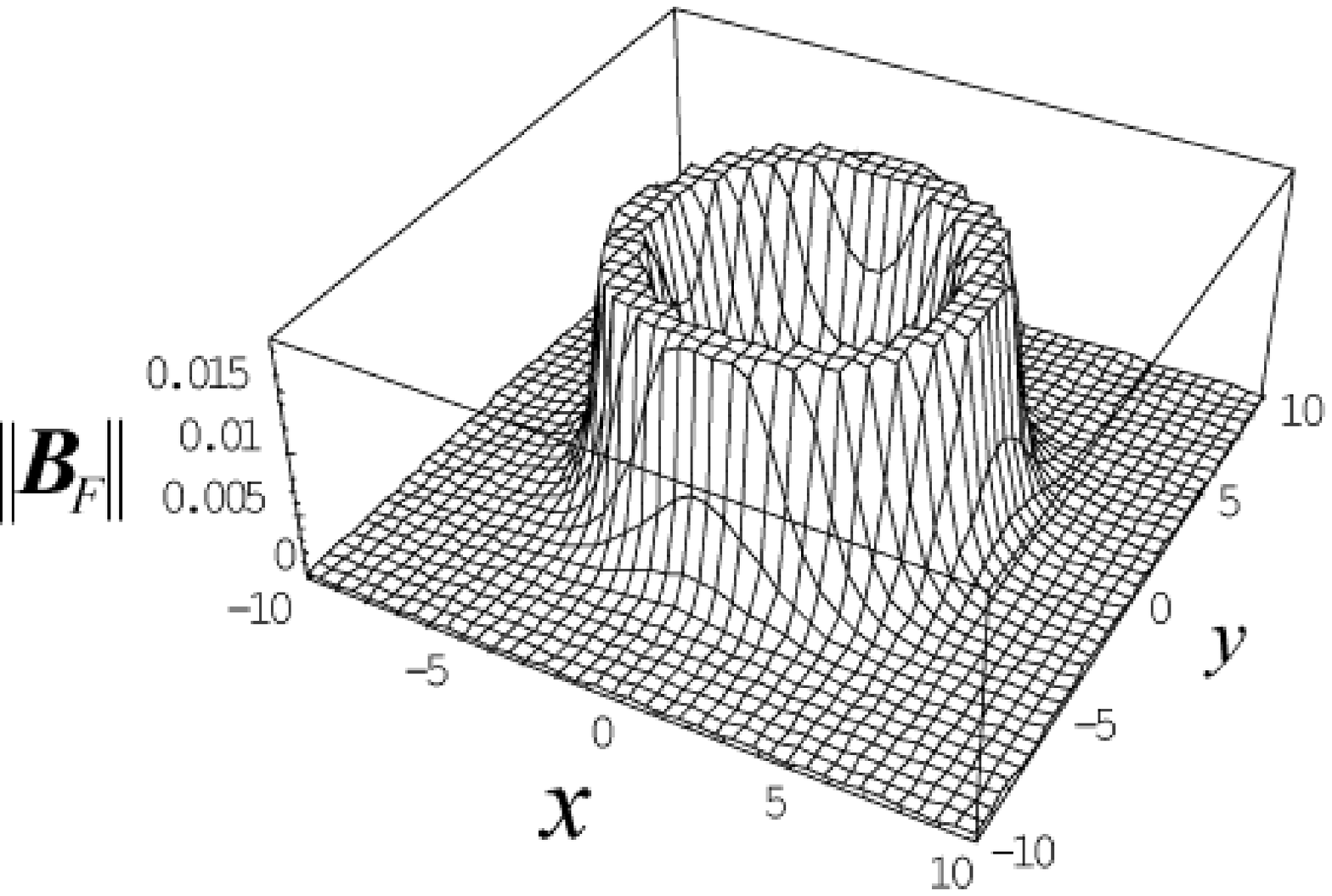}}
      \caption{Behavior of $||\bm{B}_F||$\hspace{0.8mm}　on the plane $z=1$.}
\end{figure}

\begin{figure}[htb]
  \subfigure[
t=3]{
      \centering
      \includegraphics[width=4cm]{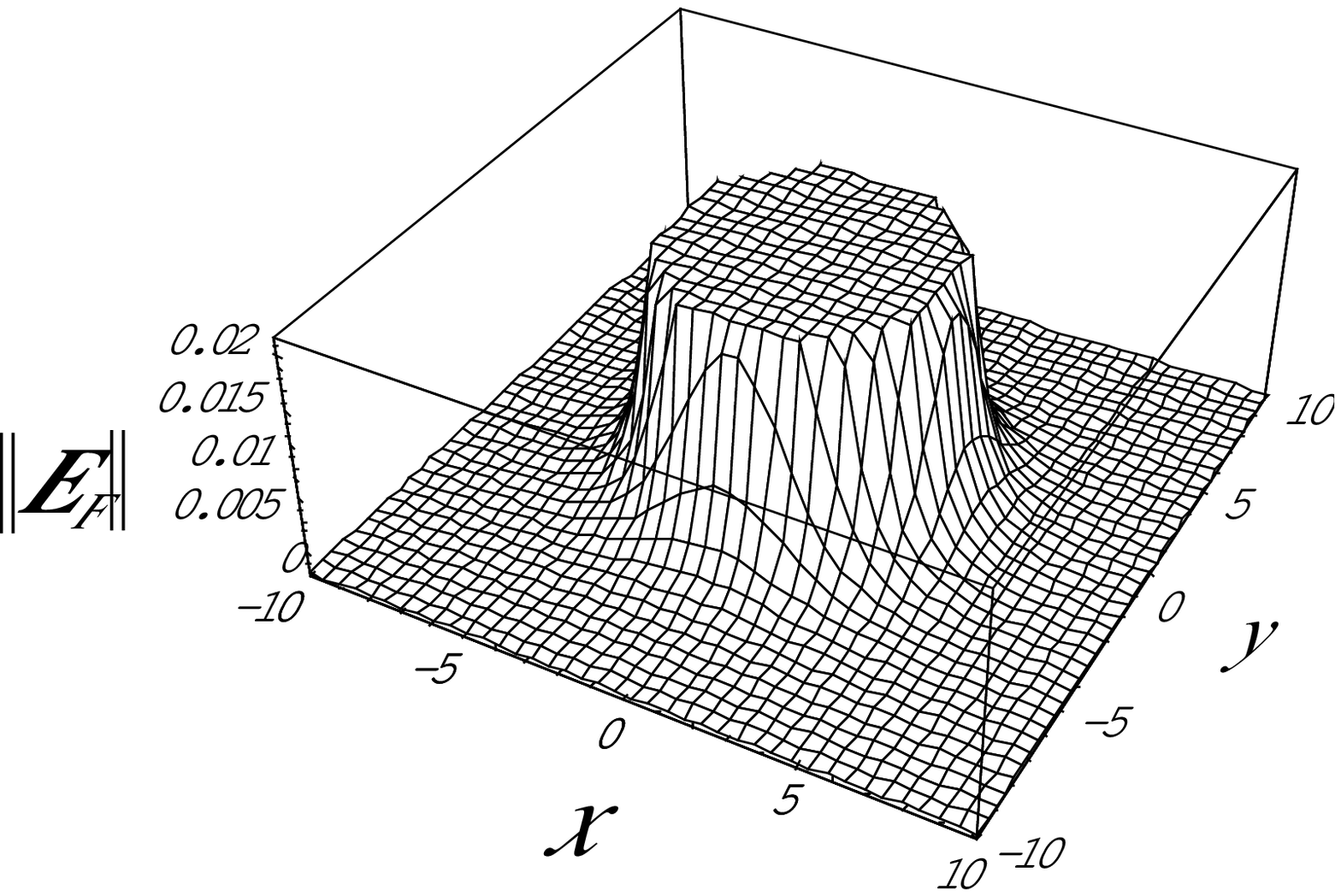}}
  \subfigure[
t=4]{
      \centering
      \includegraphics[width=4cm]{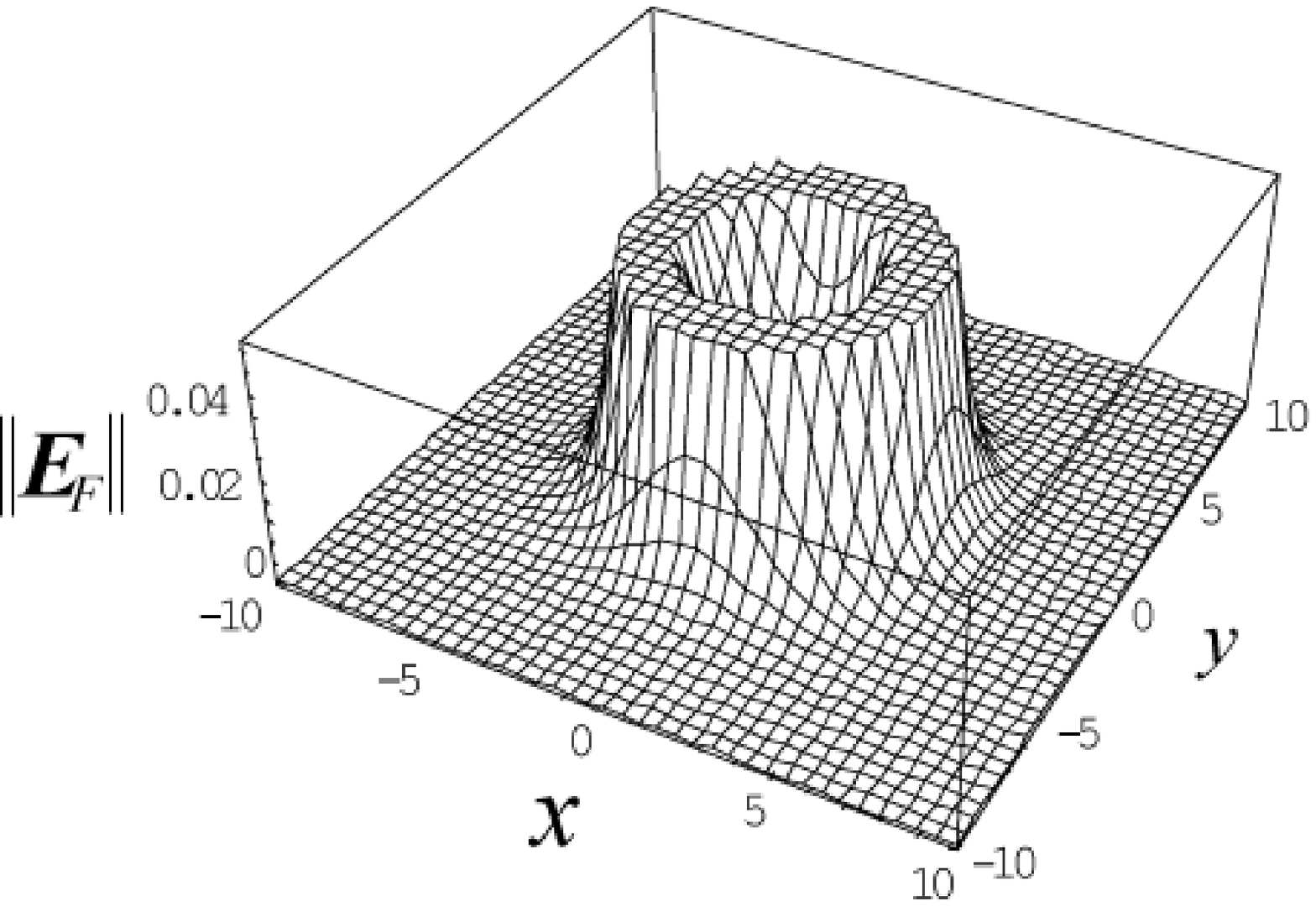}}
       \subfigure[t=5]{
      \centering
      \includegraphics[width=4cm]{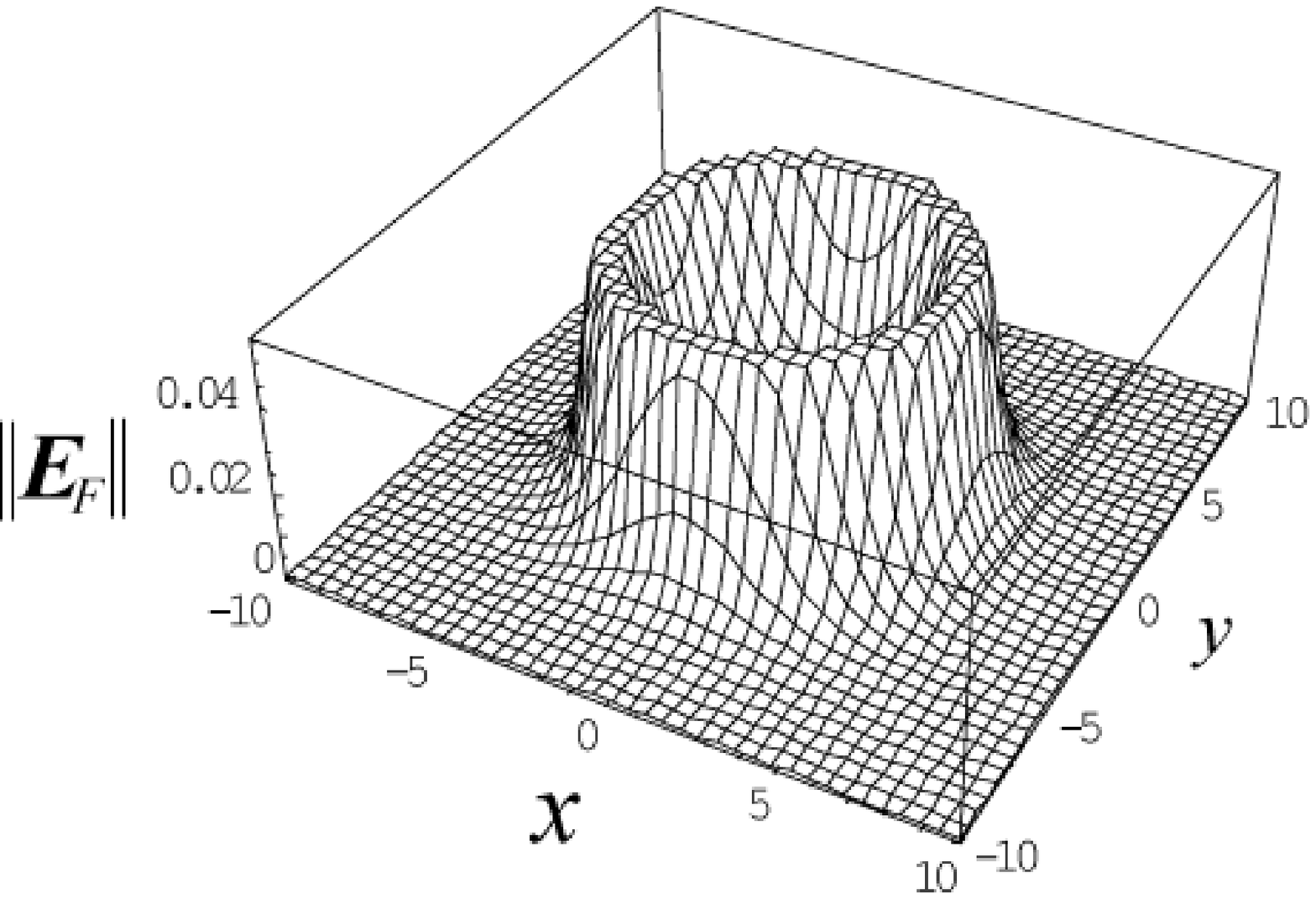}}
      \caption{Behavior of $||\bm{E}_F||$　\hspace{0.8mm} on the plane $z=1$.}
\end{figure}

\begin{figure}
  \subfigure[
t=3]{ \centering
      \includegraphics[width=4cm]{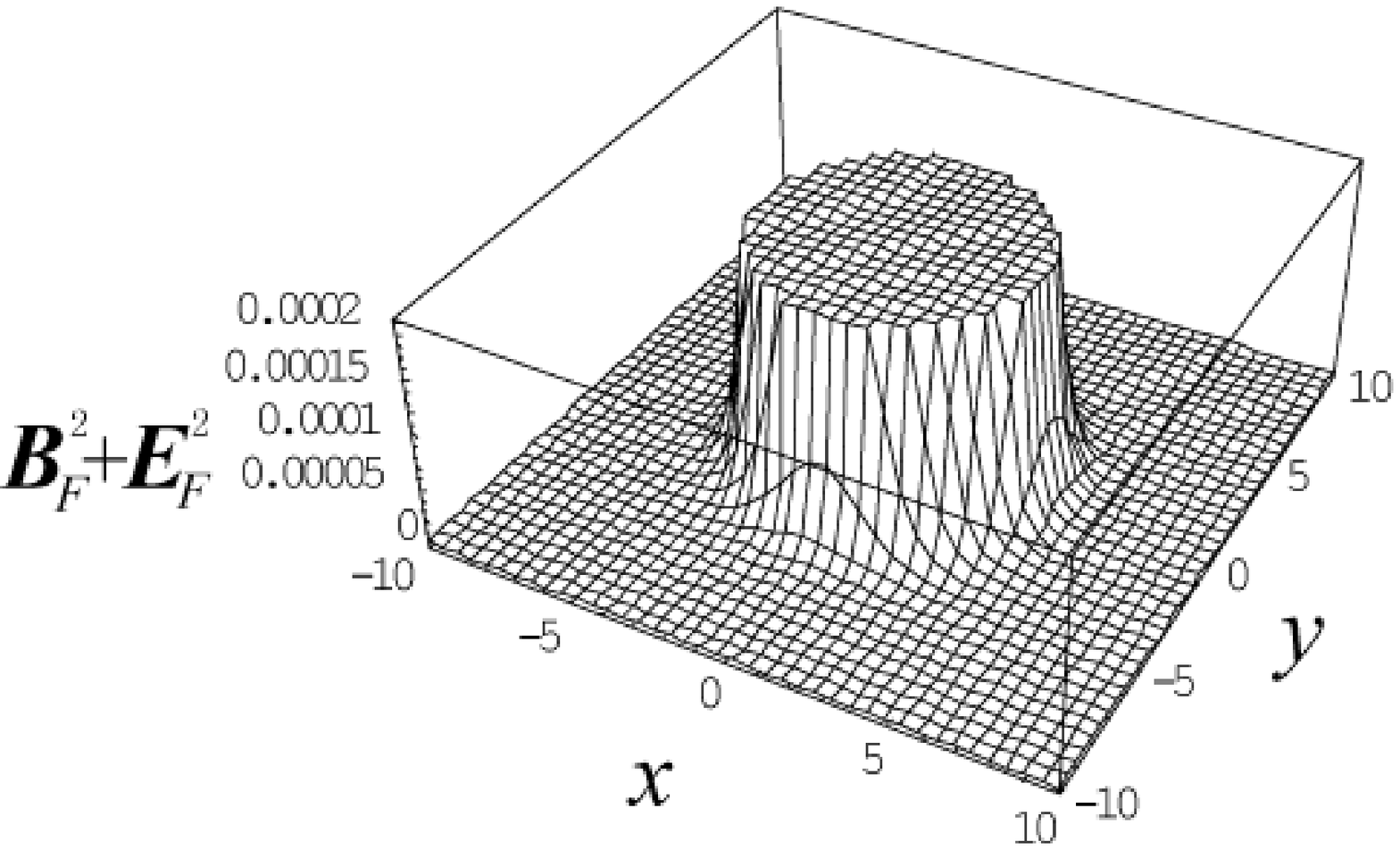}}
  \subfigure[
t=4]{
      \centering
      \includegraphics[width=4cm]{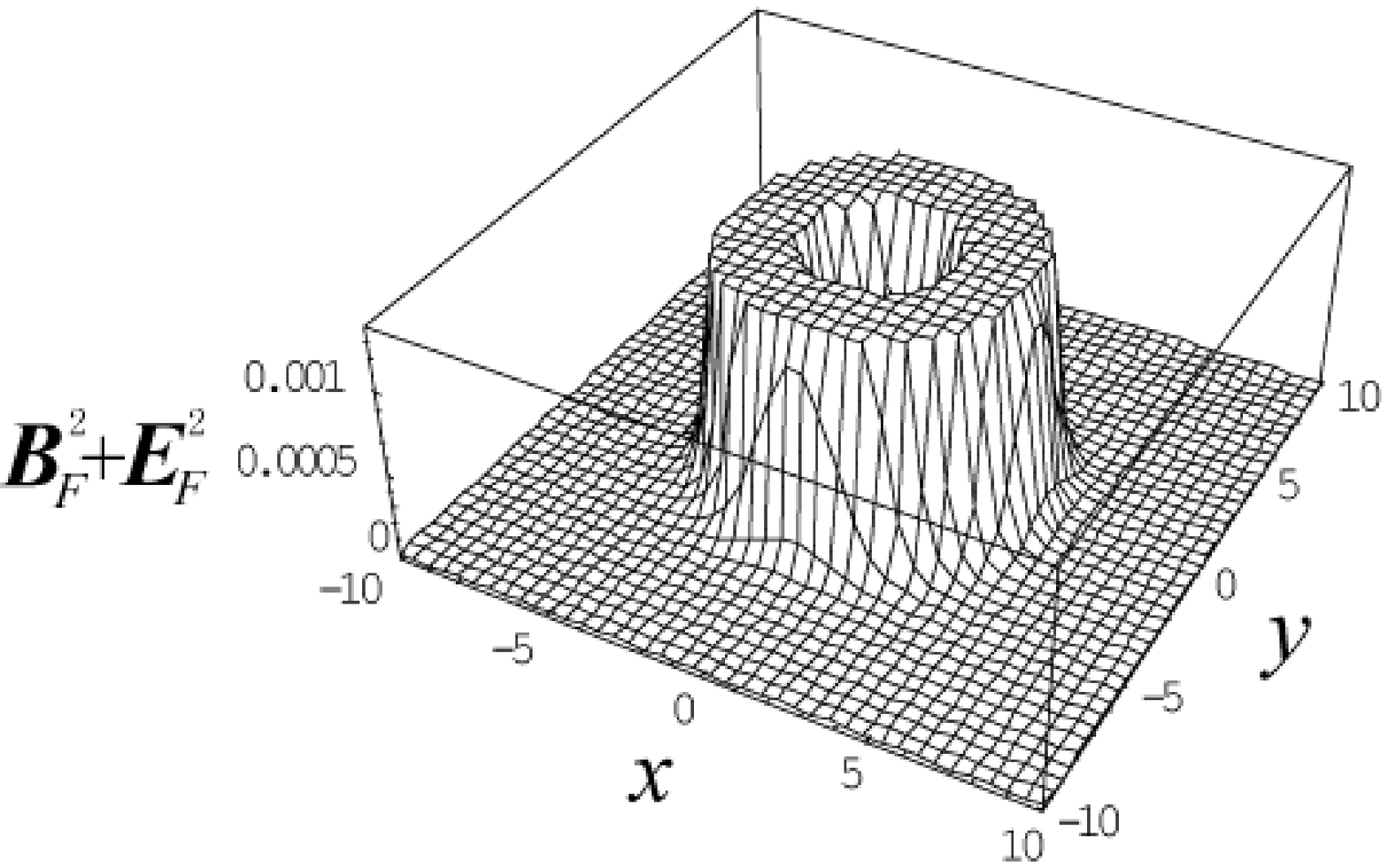}}
       \subfigure[t=5]{
      \centering
      \includegraphics[width=4cm]{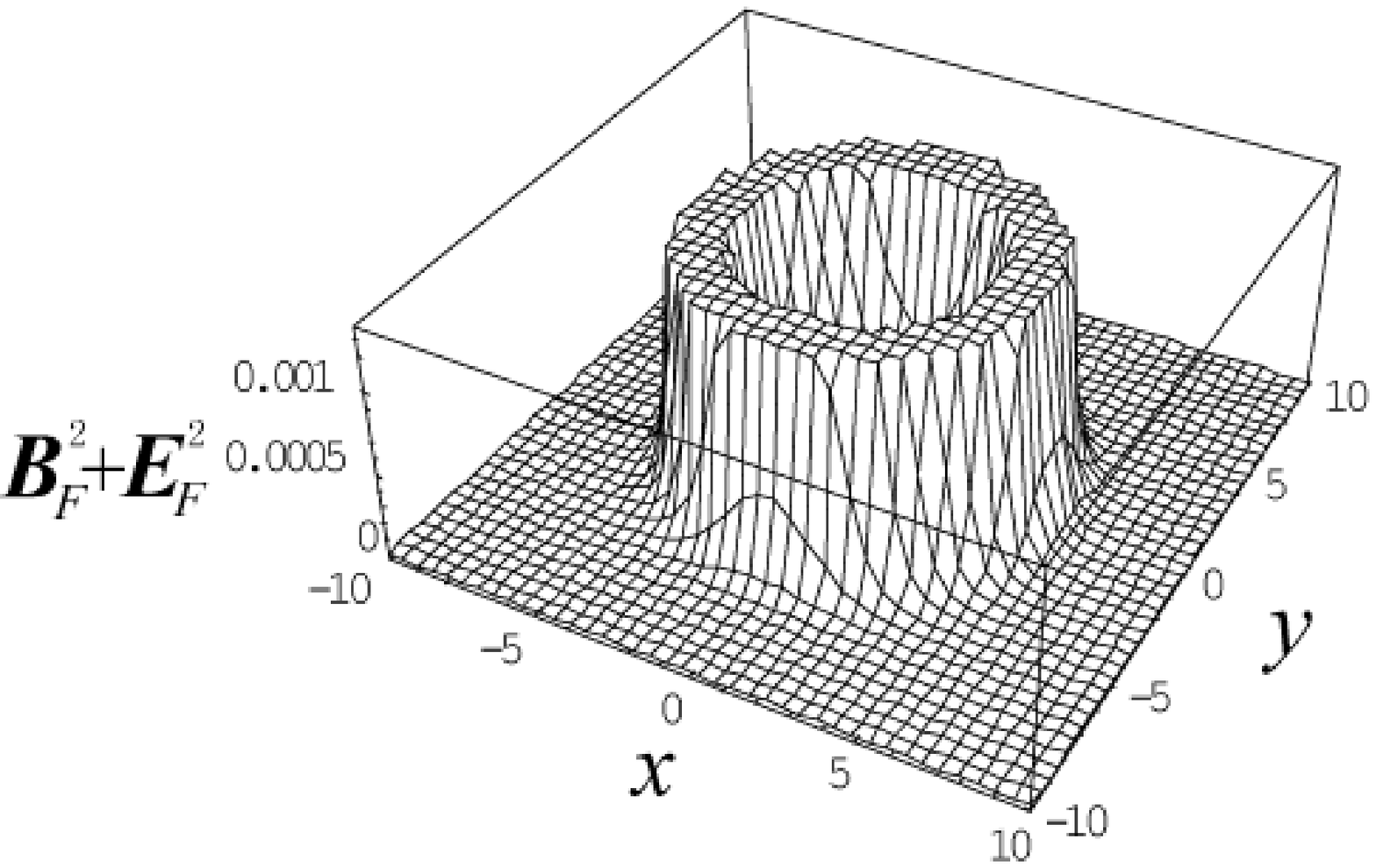}}
      \caption{Behavior of ${\bm B}_F^2+{\bm E}_F^2$ on the plane $z=1$. }
\end{figure}


 We find that these quantities are concentrated on a circle $\sqrt{x^2+y^2}$=$f(t)$
 with $f(t)$ an increasing function of $t$.\\
 The electric current density becomes
\begin{eqnarray}
&\displaystyle{j_{F, x}=\frac{-4(xz-y)}{\rho^4(Y+1)^2}},\quad
\displaystyle{j_{F, y}=\frac{-4(x+yz)}{\rho^4(Y+1)^2},}\quad
\displaystyle{j_{F, z}=\frac{-2(1+z^2+t^2-\rho^2)}{\rho^4(Y+1)^2}}
\end{eqnarray}
and hence
\begin{eqnarray}
||\bm{j}_F||=\frac{2\sqrt{L^2-4t^2\rho^2}}{\rho^4(Y+1)^2}.
\end{eqnarray}

In Fig.4, we show the time-development of $||\bm{j}_F||$. We find a
behavior similar to those of $||\bm{B}_F||$, $||\bm{E}_F||$ and
$\bm{B}_F^2+\bm{E}_F^2$.


\begin{figure}
  \subfigure[
t=3]{
      \centering
      \includegraphics[width=4cm]{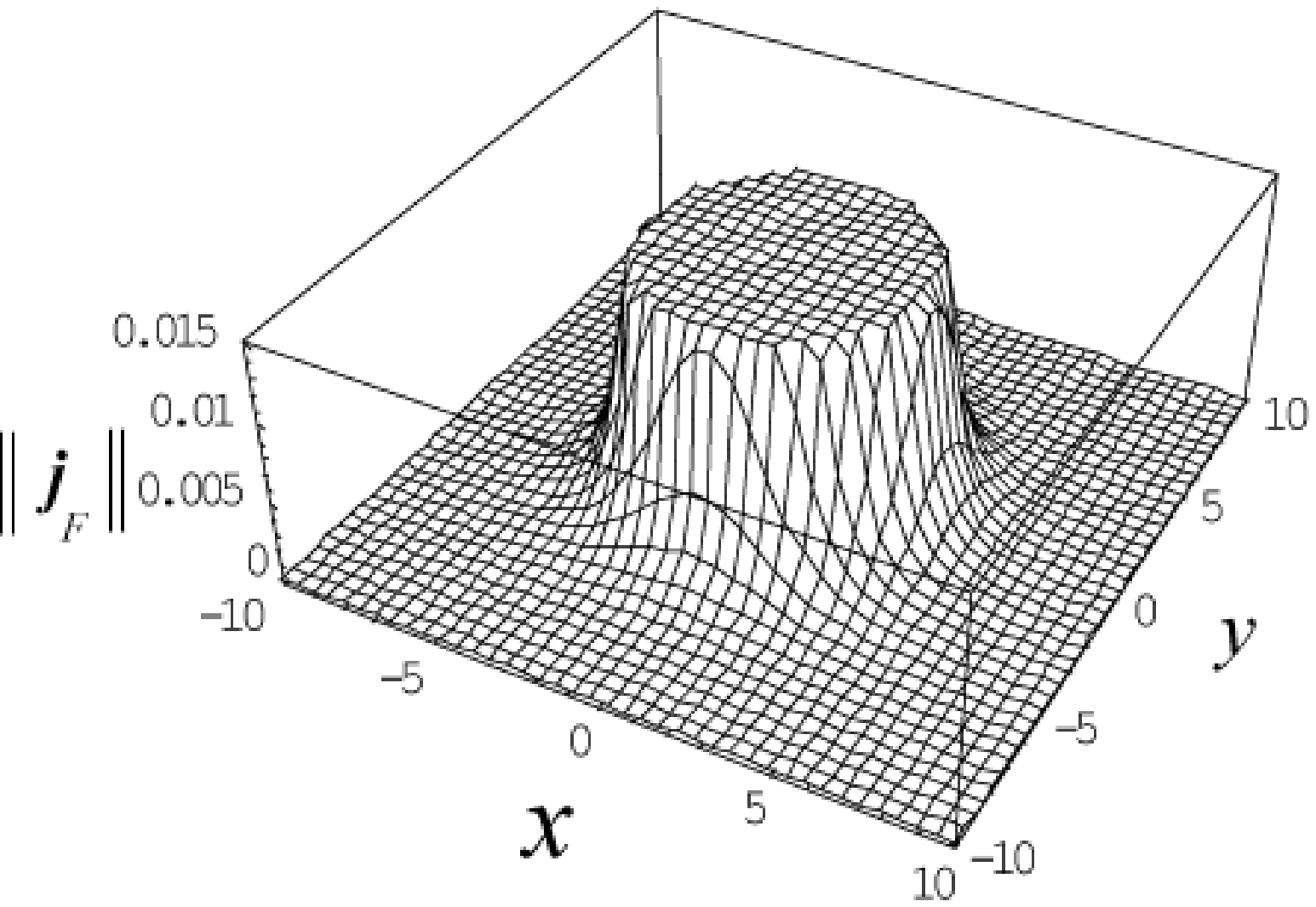}}
  \subfigure[
t=4]{
      \centering
      \includegraphics[width=4cm]{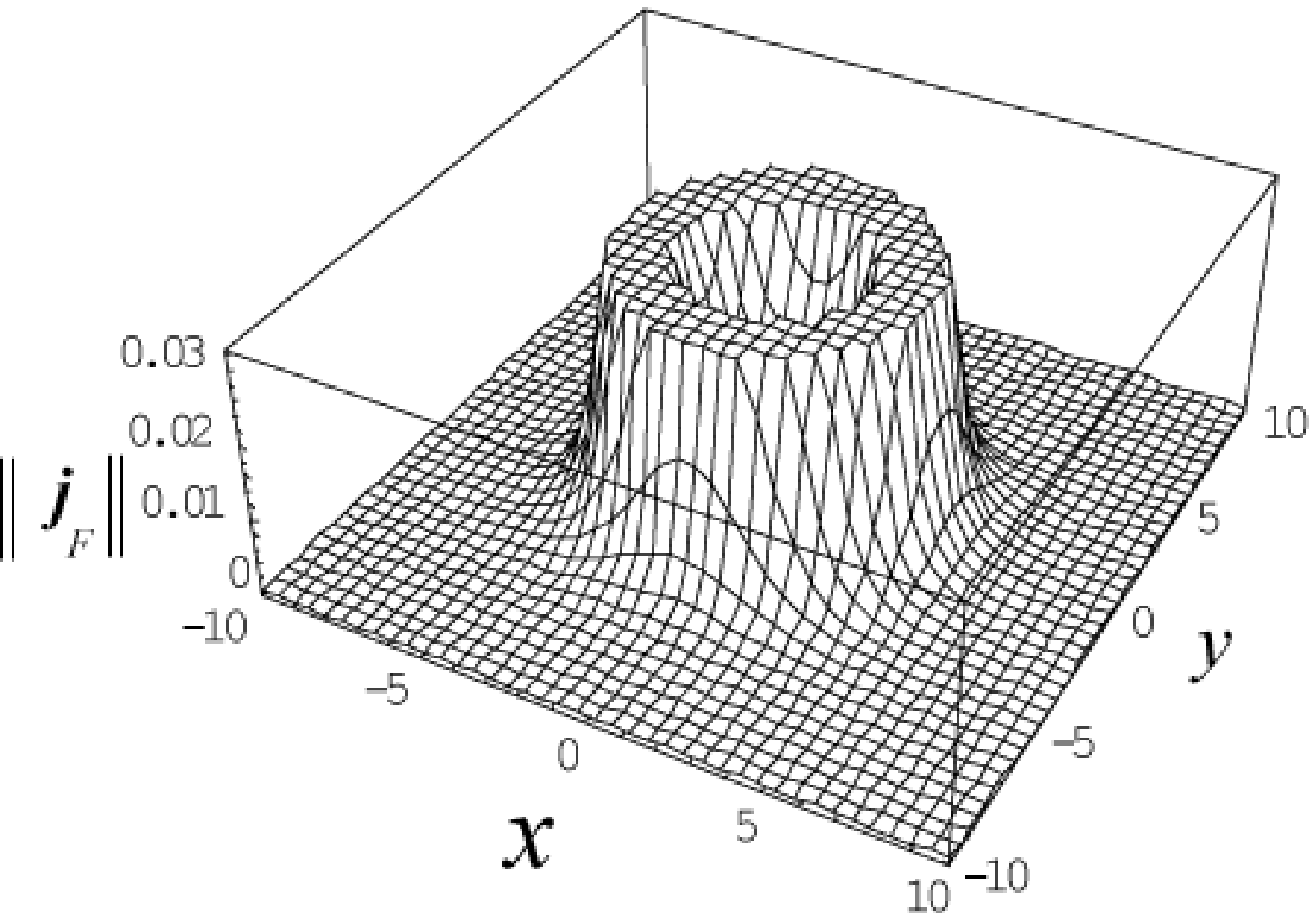}}
       \subfigure[ t=5]{
      \centering
      \includegraphics[width=4cm]{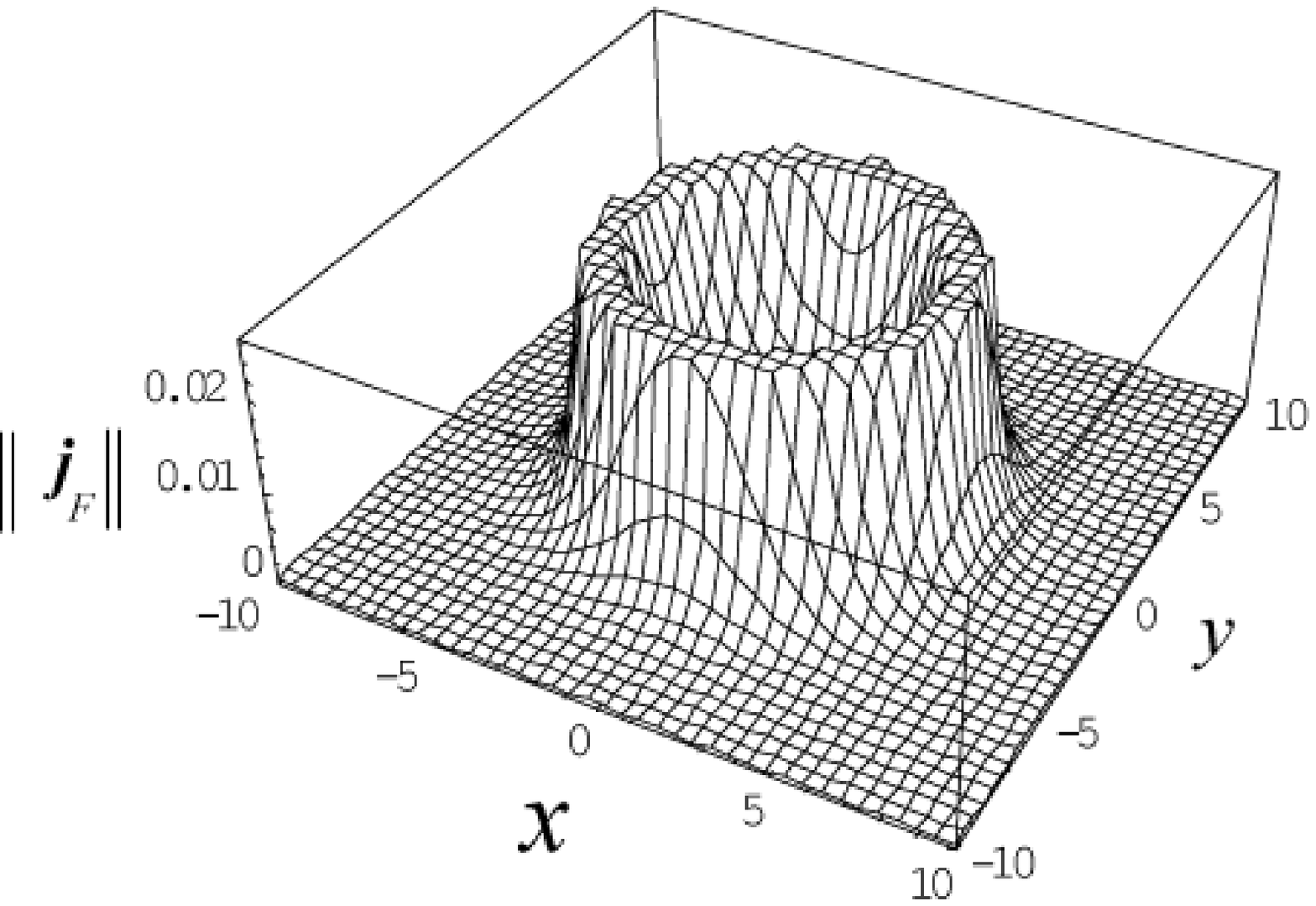}}
      \caption{Behavior of $||\bm{j}_F||$ \hspace{0.8mm}　on the plane $z=1$.}
\end{figure}


The electric charge density $\rho_F$ becomes
\begin{eqnarray}
&\displaystyle{\rho_F=\frac{-4tz}{\rho^4(Y+1)^2}}=\frac{-64tz}{\left[(1+t^2-r^2)^2+4r^2  \right]^2}.
\end{eqnarray}
In Fig.5, we show the time-development of $\rho_F$ on the plane
$z=1$.


\begin{figure}
  \subfigure[
t=1]{
      \centering
      \includegraphics[width=4cm]{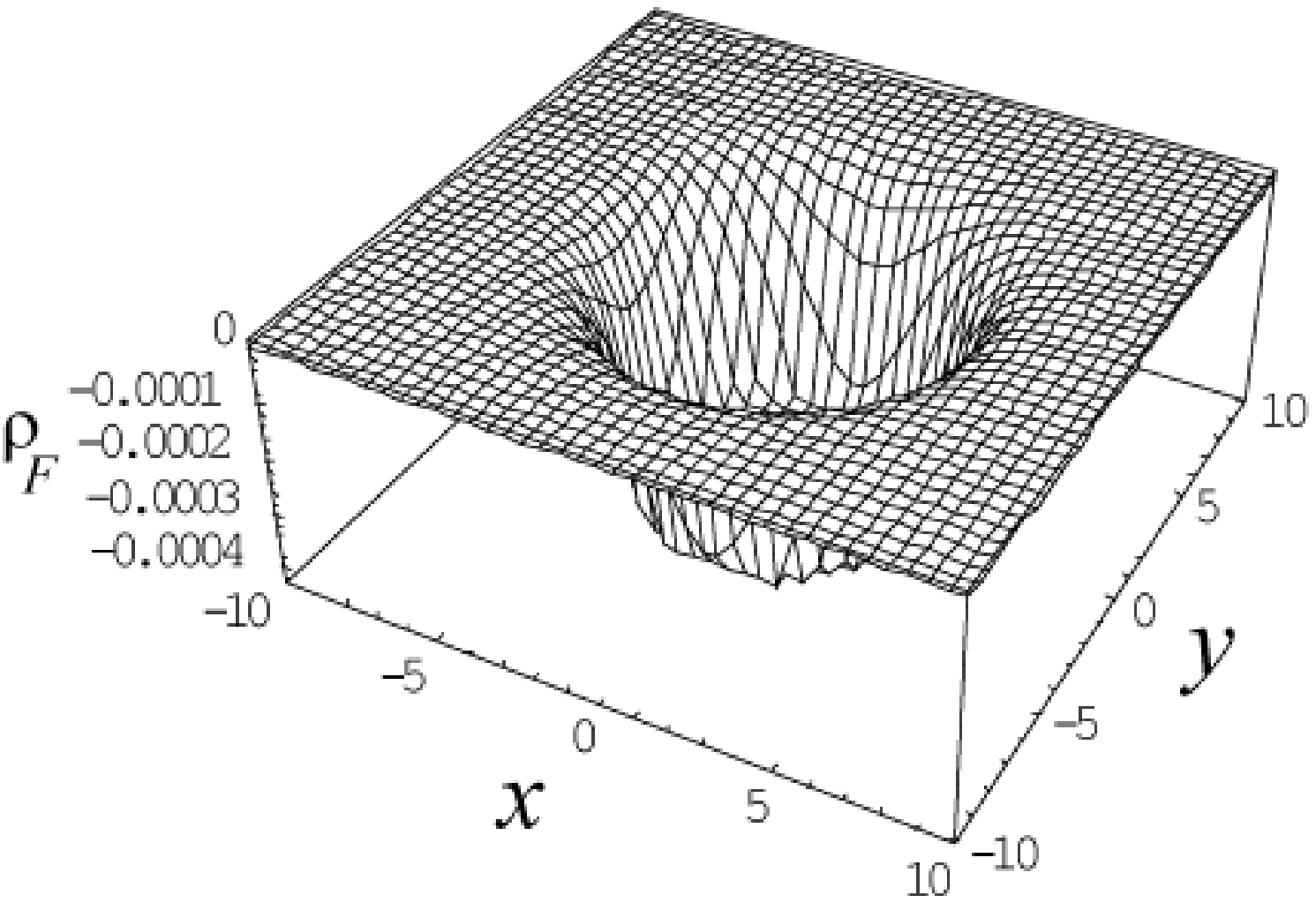}}
  \subfigure[
t=2]{
      \centering
      \includegraphics[width=4cm]{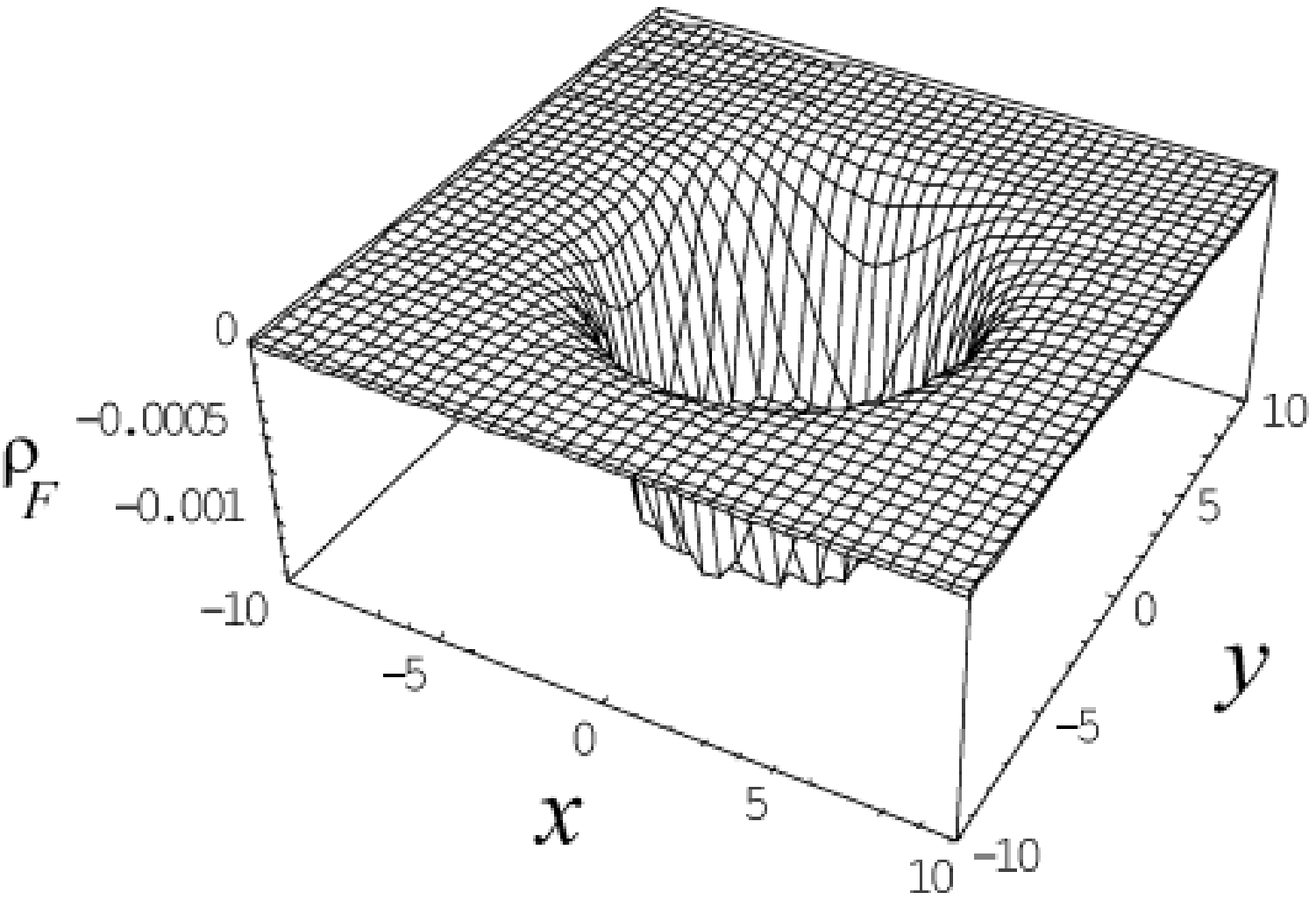}}
       \subfigure[t=3]{
      \centering
      \includegraphics[width=4cm]{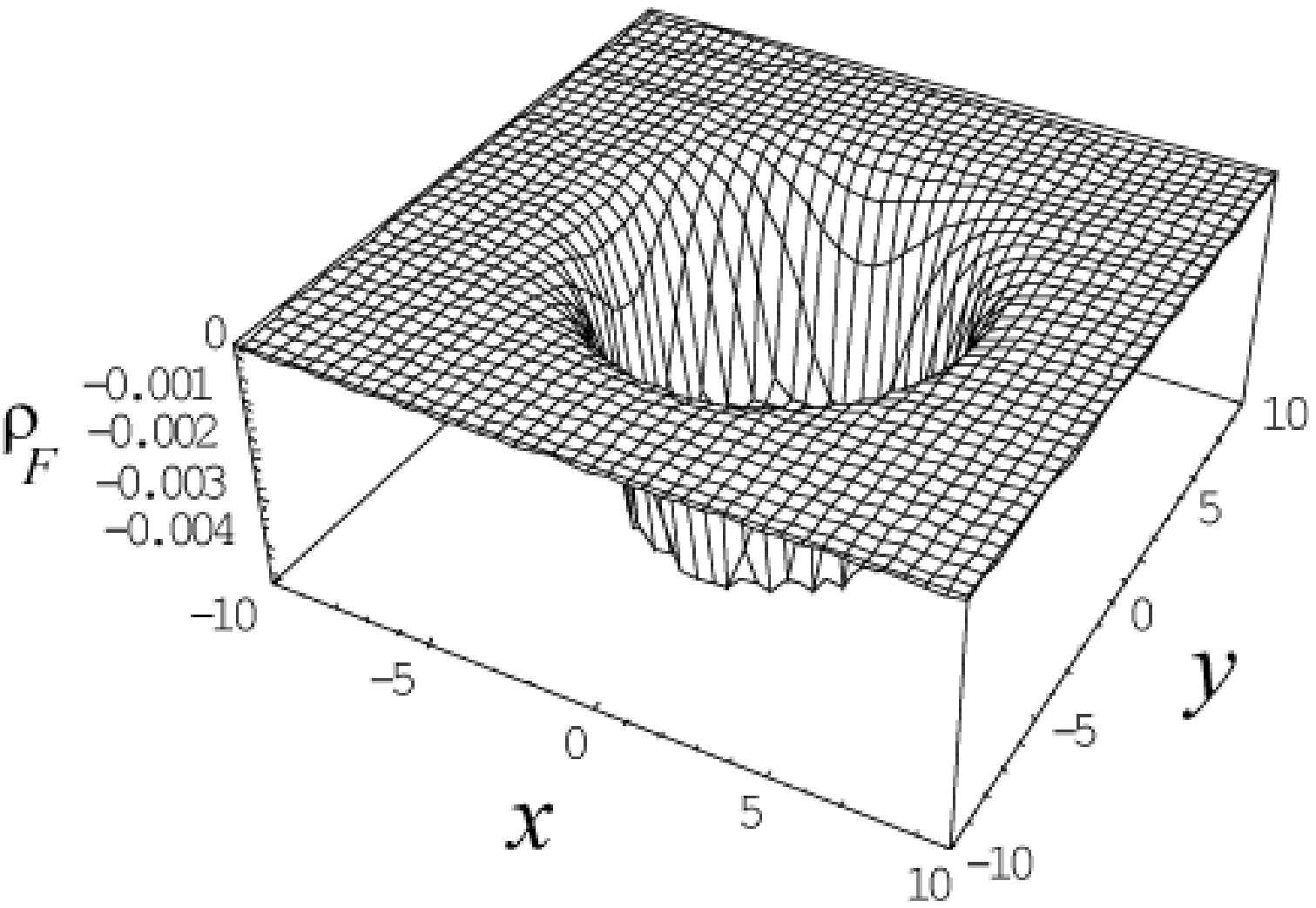}}
 \subfigure[
t=4]{
      \centering
      \includegraphics[width=4cm]{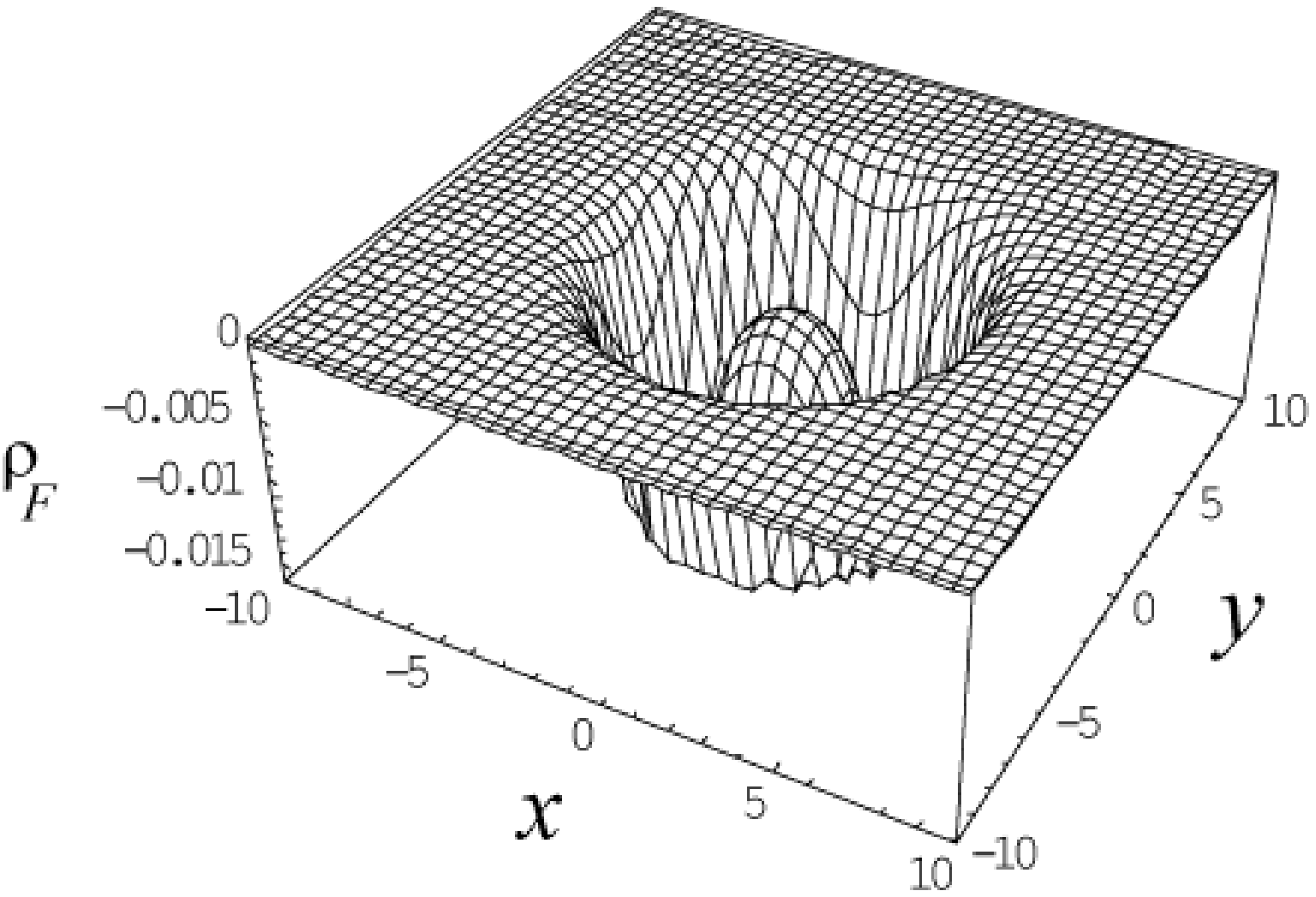}}
     \subfigure[
t=5]{
      \centering
      \includegraphics[width=4cm]{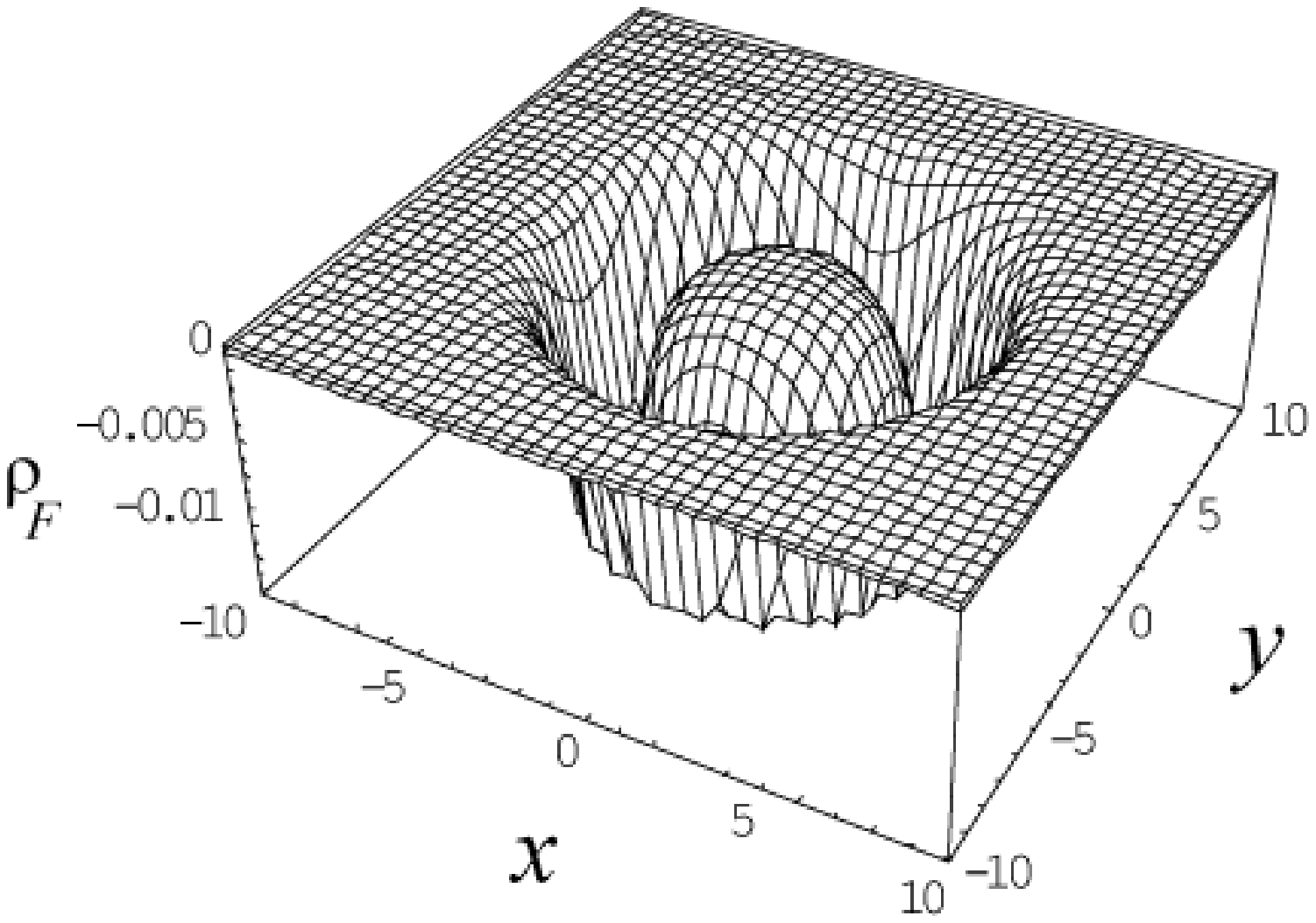}}
 \subfigure[
t=6]{
      \centering
      \includegraphics[width=4cm]{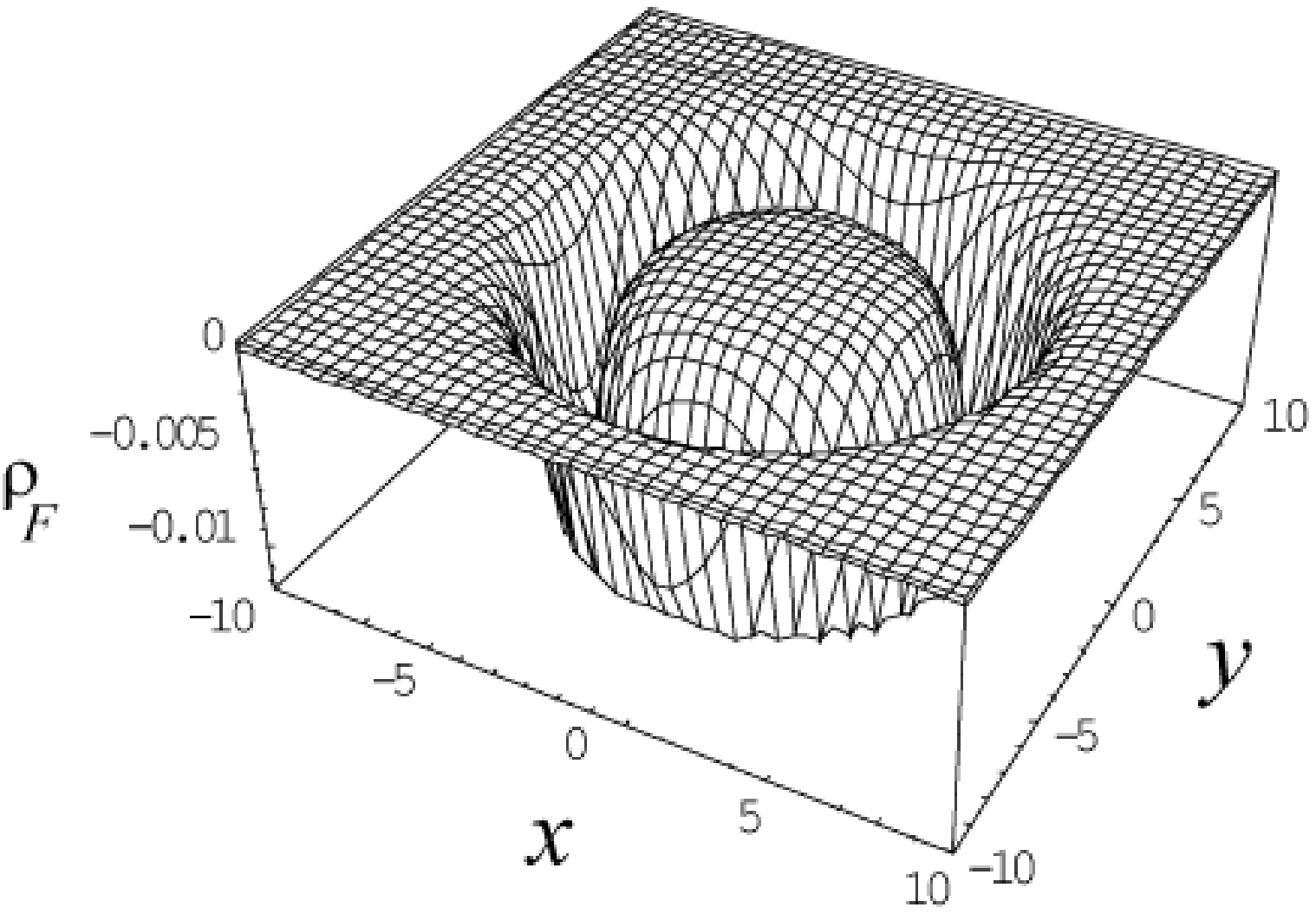}}
 \subfigure[
t=7]{
      \centering
      \includegraphics[width=4cm]{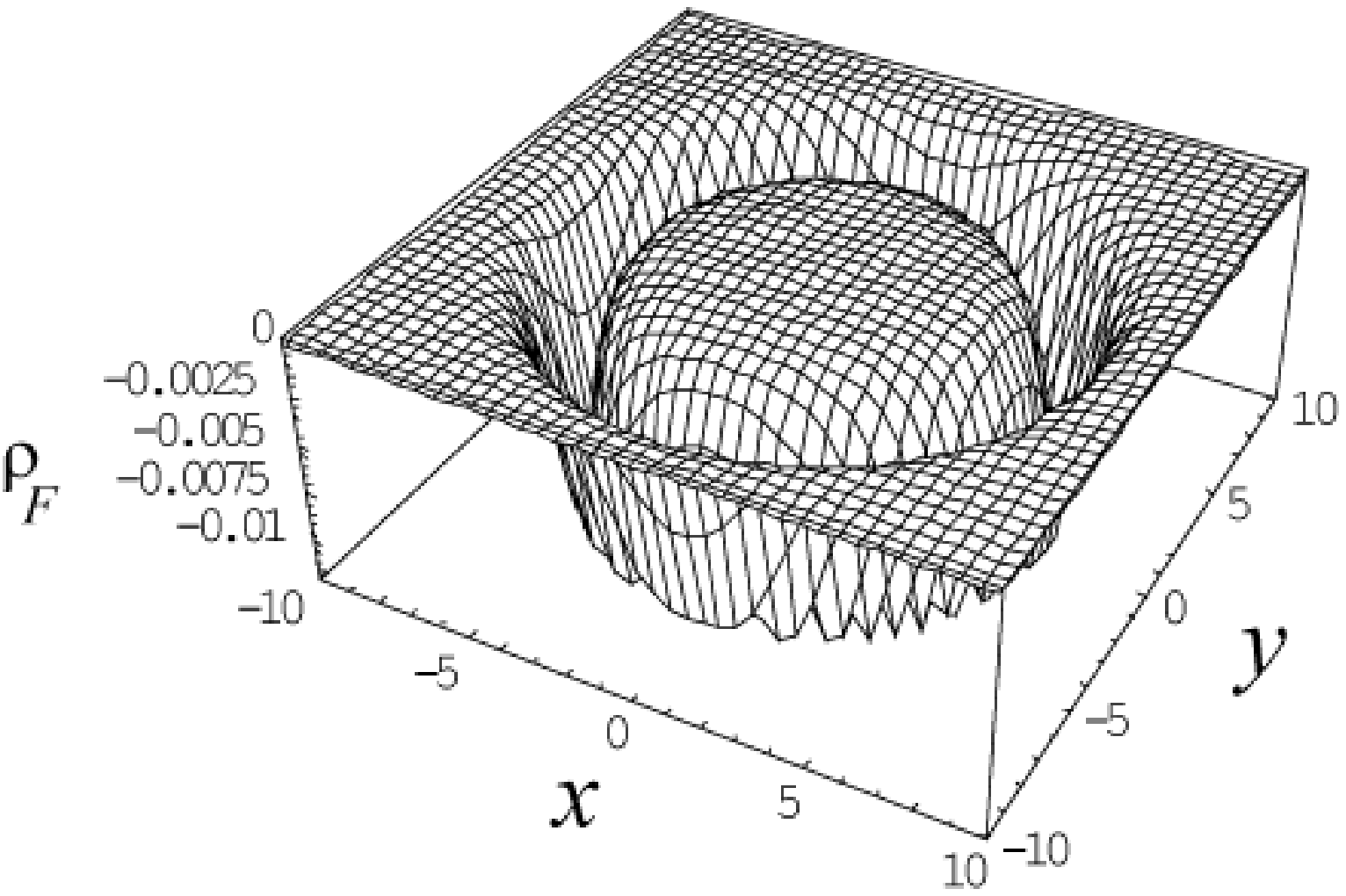}}
 \subfigure[
t=8]{
      \centering
      \includegraphics[width=4cm]{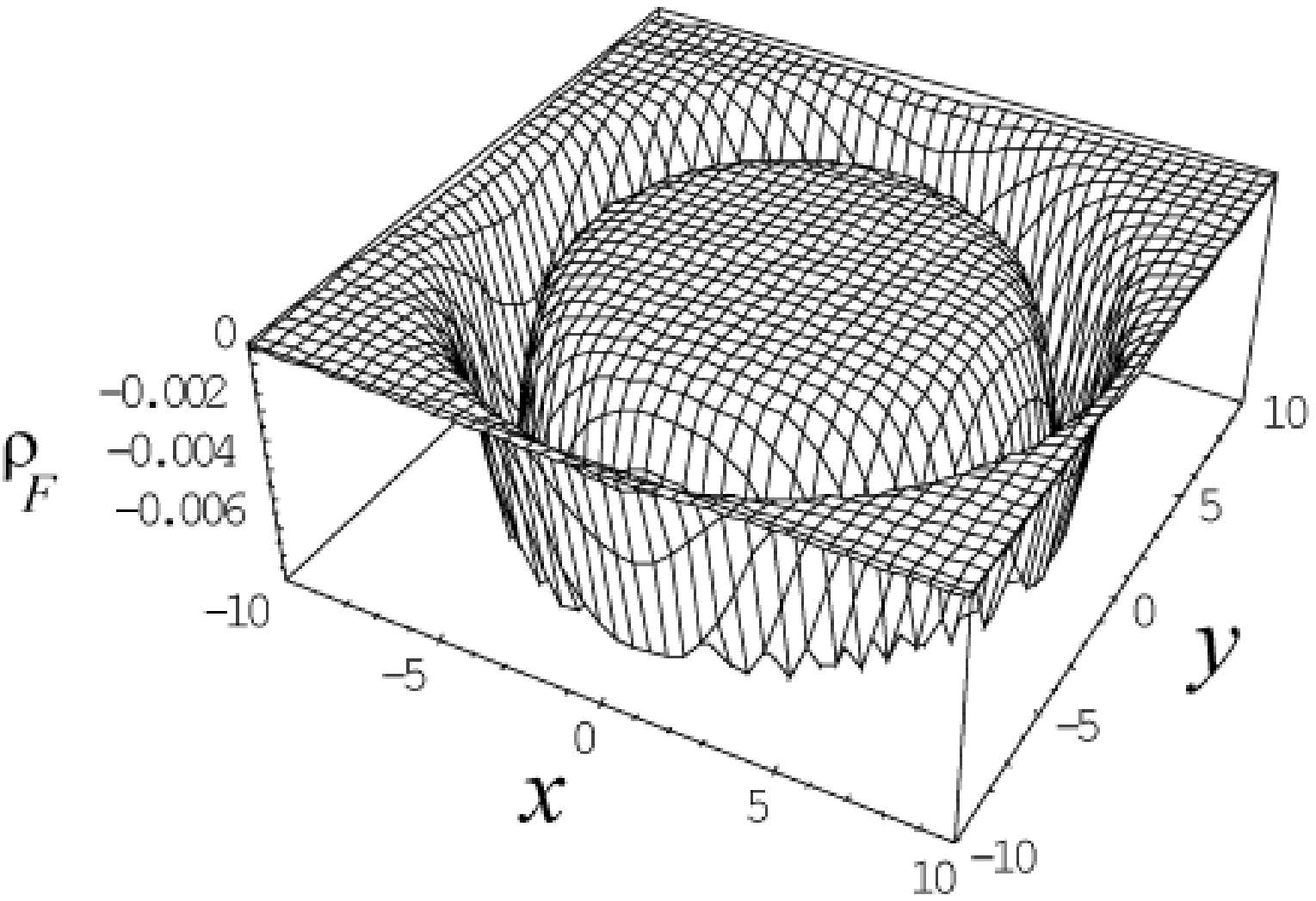}}
 \subfigure[
t=9]{
      \centering
      \includegraphics[width=4cm]{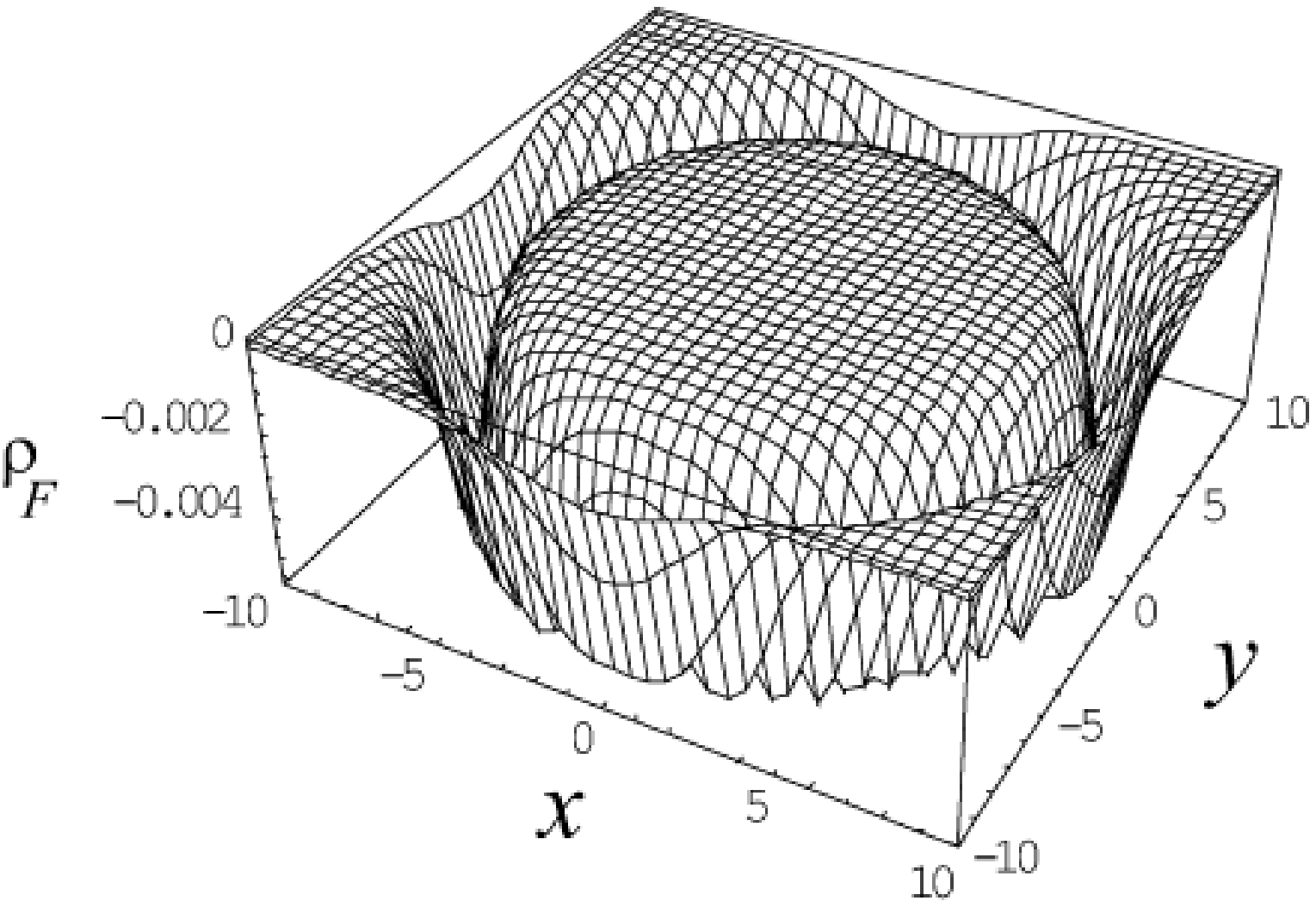}}
       \subfigure[
t=10]{
      \centering
      \includegraphics[width=4cm]{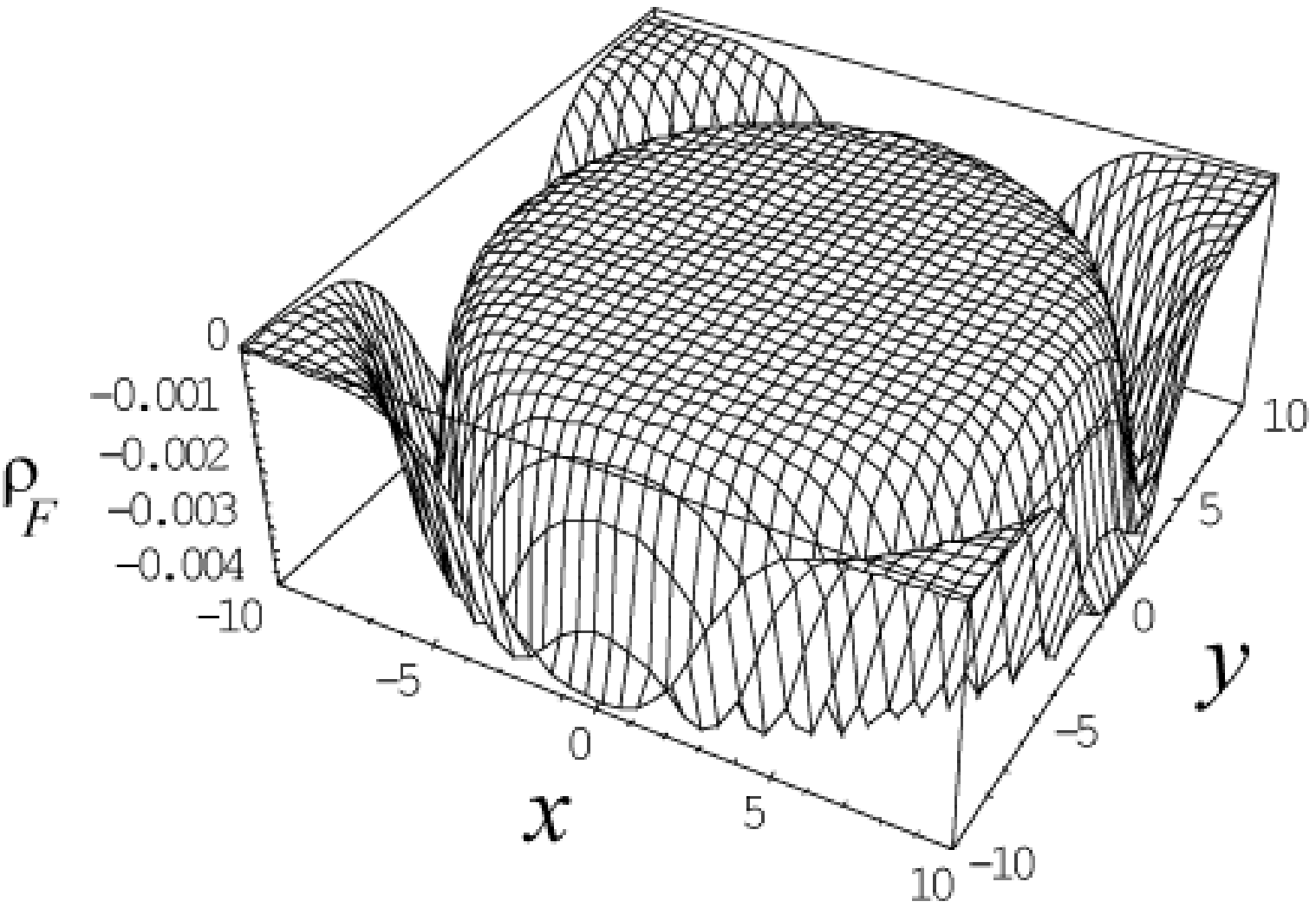}}
      \caption{Behavior of $\rho_F$　\hspace{0.8mm}　on the plane $z=1$.}
\end{figure}

On the other hand, defining $\varepsilon$ by
\begin{eqnarray}
&\displaystyle{\varepsilon\equiv\frac{1}{2}\int\!\!\!\int\!\!\!\int dxdydz \left(\bm{E}^2+\bm{B}^2 \right)\label{Energy}},
\end{eqnarray}
and observing
\begin{eqnarray}
&\displaystyle{\bm{E}^2 =\bm{B}^2
=\frac{64\left( 1+t^2+r^2+2ty \right)^2}{\left[ t^4-2t^2(r^2-1)+(r^2+1)^2 \right]^3}},
\end{eqnarray}
we obtain \begin{eqnarray}
\varepsilon_F=\frac{1}{2}\varepsilon=4\pi^2
\end{eqnarray}
from (\ref{Fer.En}), (\ref{Fer.en}).
This result is expected from the observation
\begin{eqnarray}
\boldsymbol{B}_F^2=\boldsymbol{B}^2=\boldsymbol{E}^2,\quad \boldsymbol{E}_F=0 \quad {\rm{at}}\quad t=0.
\end{eqnarray}

At $t=0$, $\rho_F$ vanishes and $\bm{j}_F$ satisfies the relation
\begin{eqnarray}
\bm{j}_F=\frac{8\boldsymbol{B}_F}{r^2+1}.
\end{eqnarray}
$||\bm{j}_F||$ simplifies to the spherically symmetric configuration
$\frac{32}{(r^2+1)^3}$. The net current defined by
\begin{eqnarray}
\bm{J}_F
=\int\!\!\!\int\!\!\!\int dxdydz \bm{j}_F
\end{eqnarray}
is non-vanishing and is given by $\left(0, 0, -\frac{8\pi^2}{3} \right)$.
\section{Summary}
We have seen that Ferreira's solution of CNLSM
defined by (1) gives rize the electromagnetic fields
$\boldsymbol{E}_F$ and $\boldsymbol{B}_F$ for
the charge density $\rho_F$ and the current density $\boldsymbol{j}_F$
satisfying the constraints (\ref{Fer.Con}).
We have investigated some of their properties.
We have shown that  we can construct exact solutions of the CNLSM
 from the examples considered in the theory of electromagnetic knots.
 We finally note that
 the fields $\bm{B}_F$ and $\bm{E}_F$ are described by a simple $4$-potential \cite{Fer}
\begin{eqnarray}
A_{F, \mu}=\frac{1}{2} \left[( g(Y)-1)(m_1 \partial_{\mu}\xi+m_3\partial_{\mu}\zeta)+m_2g(Y)\partial_{\mu} \varphi \right],
\end{eqnarray}
while the $4$-potential $A_{\mu}$ realising $\bm{B}_F$ and $\bm{E}_F$ is somewhat complicated \cite{IB}.
The knot structure of $\boldsymbol{B}_F$ is inherited  by the Hopf index associated with
$u$. On the other hand, the knot structures of $\boldsymbol{B}$ and $\boldsymbol{E}$ are inherited  by the Hopf indices associated with $\eta_m$ and $\eta_e$, respectively.
\section*{Acknowledgments}
This research was partially supported by the National Natural
Science Foundation of China (Grant No.10601031) and the Innovation
Program of Shanghai Municipal Education Commission(Grant No.
09ZZ183). One of the authors (M.H.) is grateful to Dr. Jun Yamashita
and Prof. Tetsuji Kawabe
 for communications.


\end{document}